\documentclass[12pt]{article}
\pdfoutput=1

\usepackage{epsf,epsfig,epstopdf}
\usepackage{graphicx}
\usepackage{amsmath,amssymb}
\usepackage{url}
\usepackage{slashed}
\usepackage{a4wide}
\usepackage{braket}

\usepackage[colorlinks=true, citecolor=blue, linkcolor=black, urlcolor=blue ]{hyperref}
\usepackage{cite}

\newcommand{\varotimes}{\otimes}  

\def\slash#1{#1 \hskip-0.45em /}

\newcommand{\Eqn}[1]{Eq.~(\ref{#1})}
\newcommand{\Eqns}[2]{Eqs.~(\ref{#1}) and (\ref{#2})}

\begin{document}

\thispagestyle{empty}

\begin{flushright}
{\small
DESY 14-104 \\
PSI-PR-14-04\\
ZU-TH 21/14 \\
[0.2cm]
\today}
\end{flushright}

\vspace{\baselineskip}

\begin{center}

\vspace{0.5\baselineskip} \textbf{\Large\boldmath
  Renormalization-group improved fully differential\\[8pt] cross sections for
  top pair production } \\
\vspace{2\baselineskip}
{\sc A.~Broggio$^a$, A.S.~Papanastasiou$^b$, A.~Signer$^{a,c}$}\\
\vspace{0.5cm}
{\sl ${}^a$Paul Scherrer Institut,\\
CH-5232 Villigen PSI, Switzerland\\
\vspace{0.2cm}
${}^b$DESY, Deutsches Elektronen-Synchrotron,\\ Notkestra{\ss}e 85,
D-22607 Hamburg, Germany\\ 
\vspace{0.2cm}
${}^c$Physik-Institut, Universit\"at Z\"urich,\\
Winterthurerstrasse 190, CH-8057 Z\"urich, Switzerland\\
}

\vspace*{1.0cm}

\textbf{Abstract}\\

\vspace{1\baselineskip}

\parbox{0.9\textwidth}{ We extend approximate next-to-next-to-leading
  order results for top-pair production to include the semi-leptonic
  decays of top quarks in the narrow-width approximation. The new
  hard-scattering kernels are implemented in a fully differential
  parton-level Monte Carlo that allows for the study of any IR-safe
  observable constructed from the momenta of the decay products of the
  top. Our best predictions are given by approximate NNLO corrections
  in the production matched to a fixed order calculation with NLO
  corrections in both the production and decay subprocesses.  Being
  fully differential enables us to make comparisons between
  approximate results derived via different (PIM and 1PI) kinematics
  for arbitrary distributions.  These comparisons reveal that the
  renormalization-group framework, from which the approximate results
  are derived, is rather robust in the sense that applying a realistic
  error estimate allows us to obtain a reliable prediction with a reduced
  theoretical error for generic observables and analysis cuts. }

\end{center}

\newpage

\section{Introduction}

The top quark, as the heaviest known fundamental particle, plays a
special role in the Standard Model (SM). Detailed investigations of
its mass, width and couplings provide crucial input for consistency
checks of the SM and in searches for physics beyond the SM. The main
production mechanism for top quarks at the Tevatron and Large Hadron
Collider (LHC) is the simultaneous production of a top anti-top quark
pair.  At leading order (LO), i.e. at ${\cal O}(\alpha_s^2)$ in an
expansion in the strong coupling $\alpha_s$, two partonic channels
contribute, $q\bar{q}$ and $gg$, with the latter (former) dominating
at the LHC (Tevatron). Beyond LO, additional
partonic channels such as $gq$ must be taken into account.

Due to its huge significance it is imperative to obtain the best
possible theoretical description of $t\bar{t}$ production processes
and there is extensive literature on this subject. The first
calculation at next-to-leading (NLO) order in QCD, i.e at ${\cal
  O}(\alpha_s^3)$, was done for the total cross
section~\cite{Nason:1987xz}, followed by the computation of the NLO
corrections to the differential cross section~\cite{Beenakker:1988bq,
  Nason:1989zy, Beenakker:1990maa}, treating the top quarks as stable
external particles. Later, the impact of electroweak corrections to
stable top quark pair production was also
investigated~\cite{Beenakker:1993yr, Kuhn:2005it, Bernreuther:2005is,
  Kuhn:2006vh, Bernreuther:2006vg, Moretti:2006nf}.

Of course, in experiments it is not the top quark itself that is
detected, but rather its decay products. In the SM the
top decays almost exclusively to a $W\, b$ pair, with the $b$ quark
resulting in a $b$-jet, $J_b$, and the $W$ decaying further, either
leptonically or hadronically. Thus, in an experimental analysis, top
quarks are reconstructed from their decay products after suitable cuts
are applied.

Since the width of the top quark is considerably larger than
$\Lambda_{\rm QCD}$ its decay can be described reliably using
perturbation theory. Furthermore, the decay products carry the
information of the spin of the parent top, resulting in so-called
spin-correlation effects. In order to allow for a more realistic
comparison with experimental data, fixed-order calculations of stable $t\bar{t}$
production have been first improved by adding the decay of the on-shell top
quarks in the narrow-width approximation (NWA). NLO corrections have been 
included both to the production as well as to the decay of the top 
quarks~\cite{Bernreuther:2004jv, Melnikov:2009dn,
  Campbell:2012uf}, while taking into account all spin-correlation
effects. 
Once the decay of the top quarks is included and they are not treated
as external particles, the (intermediate) top quarks of course need not
necessarily be on-shell. Off-shell effects in $t\, \bar{t}$ production
have been studied in two independent NLO calculations of $W^+b\,
W^-\bar{b}$ production~\cite{Denner:2010jp, Bevilacqua:2010qb,
  Denner:2012yc} and in an effective-theory
framework~\cite{Falgari:2013gwa}. 

Ideally, to allow for a direct comparison with experimental data,
these calculations have to be matched to a parton shower. For stable $t\bar{t}$
production (at NLO) this was first achieved using the frameworks of {\sc MC@NLO} 
\cite{Frixione:2003ei} and {\sc POWHEG} \cite{Frixione:2007nw}. Here the top 
quarks are treated as stable within the LesHouches events and the top decays 
are performed by the parton shower itself. Using an unweighting procedure
\cite{Frixione:2007zp} (see also \cite{Artoisenet:2012st}) it is possible to 
include the top decay with spin-correlations and also the leading off-shell
effects at the level of Monte Carlo LesHouches events.  

Given the ever increasing precision of experimental results, it is
highly desirable to improve theoretical predictions of top quark pair
production beyond NLO.  A major breakthrough was made by the recent
computation of the total cross section (for stable, external top
quarks) at next-to-next-to-leading order (NNLO) in
QCD~\cite{Baernreuther:2012ws, Czakon:2012zr,
  Czakon:2012pz,Czakon:2013goa} which includes all terms of ${\cal
  O}(\alpha_s^4)$.  However, at the time of writing, only the total
cross section is known at this accuracy, though considerable progress
is being made to obtain predictions for more exclusive observables at
this order in perturbation theory~\cite{Bonciani:2008az,
  Bonciani:2009nb, Bonciani:2010mn, Bonciani:2013ywa, Korner:2008bn,
  Kniehl:2008fd, Bierenbaum:2011gg, Abelof:2014fza}.

An alternative approach to go beyond NLO -- that is to some extent
complementary to fixed-order NNLO calculations -- is to apply
threshold resummation and obtain renormalization-group (RG)-improved
results. In this way, logarithmically-enhanced contributions related
to the emission of soft gluons can be resummed. By re-expanding the
resummed cross section to fixed-order in perturbation theory and
assuming the logarithmically-enhanced terms are dominant, it is also
possible to obtain approximate higher-order expressions for the hard
scattering kernels.

The precise definition of the soft parameter and hence the exact form
of the logarithm to be resummed, depends on the kinematics which in
turn depends on the particular observable that is analyzed. For the
total cross section, it is possible to work in the limit of a $t
\bar{t}$ pair nearly at rest. The small parameter in this case is the
velocity of the top quarks, $\beta \equiv \sqrt{1-4m^2_t/s}$, where
$\sqrt{s}$ is the partonic centre of mass energy and $m_t$ is the mass
of the top quark. Apart from potentially large logarithms of the form
$\alpha_s^n (\ln\beta)^{m}$ (with $m\le 2 n$) relative to the Born
term, there are also Coulomb contributions of the form
$(\alpha_s/\beta)^n$ that can be resummed in this approach. This
approach has been used to compute the total cross section up to
next-to-next-to-leading logarithmic accuracy (NNLL)~\cite{Moch:2008qy,
  Czakon:2008cx, Langenfeld:2009wd, Beneke:2009rj, Czakon:2009zw,
  Beneke:2009ye, Beneke:2010da, Aliev:2010zk, Beneke:2011mq,
  Cacciari:2011hy}.

If one is interested in more detailed information on the final state
it is advantageous to consider a different `soft' kinematics that does
not require the top pair to be nearly at rest. In pair-invariant mass
(PIM) kinematics~\cite{Kidonakis:1998bk} the soft parameter is
identified by the quantity $\sqrt{s} (1-z)$, where $z = M^2/s$ and $M$
is the invariant mass of the top pair. The soft parameter in
single-particle inclusive (1PI) kinematics~\cite{Laenen:1998qw,
  Catani:1998tm} is identified by the variable $s_4=(p_{\bar{t}}+k)^2
- m^2_t$, where $p_{\bar{t}}$ is the 4-momentum of the unobserved
anti-top quark and $k$ is the 4-momentum of the real radiation in the
final state.  The partonic threshold region is defined as a limit of
the soft parameter (soft limit) which describes the emission of soft
real radiation in the scattering process.  In PIM the soft limit
corresponds to $z\to 1$ while in 1PI it corresponds to $s_4\to 0$.

Using PIM kinematics, theoretical predictions for the invariant mass
distribution at NNLL have been obtained~\cite{Ahrens:2010zv}, whereas
the transverse momentum and rapidity distributions of the (anti-)top
quark have been computed at NNLL~\cite{Kidonakis:2010dk,Ahrens:2011mw,
  Guzzi:2014wia} applying 1PI kinematics.  These results extend
previous calculations obtained both in PIM and 1PI kinematics at NLL
accuracy in \cite{Kidonakis:1996aq, Kidonakis:1997gm,
  Kidonakis:2000ui, Kidonakis:2001nj, Kidonakis:2003qe, Banfi:2004xa,
  Kidonakis:2008mu}. Recently the special case of highly boosted top
quarks has also been investigated in both
kinematics~\cite{Ferroglia:2012ku, Ferroglia:2013zwa,
  Ferroglia:2013awa}. Integrating the various differential cross
sections (or using the small $\beta$ approximation) also allows one to
obtain the total cross section at NNLL. By re-expanding these results,
approximate NNLO predictions for the total cross section were obtained
prior to the completion of the full NNLO calculation. Whatever
resummation is used, there will always be non-logarithmic terms that
are missing in these approximate results.  Since they differ by
subleading terms, comparing results obtained by the various approaches
is a useful tool to estimate the importance of the omitted
contributions and, thereby, obtain a realistic estimate of the
theoretical error. Of course, for the total cross section, the full
NNLO result is now known and can be used to validate (or not) the
procedure. Generally speaking, using a conservative estimate of the
theoretical uncertainties, approximate NNLO predictions obtained from
re-expanding RG-improved results are in reasonably good agreement with
the exact NNLO result. A similar assessment of the quality of this
method and the accompanying uncertainty can be made by comparing
approximate NLO results to the complete NLO results for various
distributions.

With this in mind, a natural next step is to generalize the
RG-improved NNLL predictions for differential cross sections to a
completely exclusive level. To this end, a fully differential NLO
partonic Monte Carlo (MC) program, including corrections to the decay
of the (anti-)top quark, can be improved, {\it{as a first step,}} by
adding approximate NNLO corrections obtained from re-expanded
RG-improved calculations.\footnote{We note that this is not an
  entirely new idea - using resummed results to improve fully
  differential results, in $e^+e^-$ collisions and in Drell-Yan
  production, is a key concept of the {\sc Geneva} MC
  \cite{Alioli:2012fc}.} In this article we present such a
calculation. We have implemented PIM and 1PI kinematics and extended
previous approximate NNLO results by adding the spin-correlated decay
of the top quarks. Comparing 1PI and PIM implementations, that differ
from each other by subleading terms, and studying a large variety of
observables at various collider setups, allows us to critically assess
the reliability of the method. Of course, such a calculation is not
meant to be a replacement for a complete fully differential NNLO
result including the decay of the top quarks. But, as we will argue,
it does allow us to improve NLO predictions at the partonic level for
a wide variety of observables including arbitrary cuts on the final
state jets, leptons and missing energy. Unlike the total inclusive
cross section, invariant-mass distributions or transverse momentum
distributions, the exclusive observables our framework allows us to
examine are much closer to the quantities actually measured in the
experiments.

Apart from its direct phenomenological use for top pair production at
hadron colliders, this work also addresses the important question on
the reliability and universality of the {\it soft gluon} resummation
procedure in obtaining approximate higher-order predictions for
generic observables. In particular, we can study for example the
transverse momentum distribution of the (anti-)top using the (`wrong')
PIM kinematics and compare to the results obtained using the
(`correct') 1PI kinematics. As we will see, the prominent features of
resummation are rather robust, giving further assurance that the
methods presented here provide us with a reasonable approximation and
an improvement compared to fixed order NLO calculations.

The work presented here relies heavily on RG-improved calculations at
NNLL using the Soft-Collinear Effective Theory (SCET) approach
presented in Refs.~\cite{Ahrens:2010zv, Ahrens:2011mw}. Hence, we will
start by presenting PIM and 1PI kinematics within the SCET framework
in Section~\ref{sec:calc} and discuss the way in which the results of
\cite{Ahrens:2010zv,Ahrens:2011mw} have to be modified in order to
include the decays of the top quarks. Here we also give details of our
MC implementation of the approximate (N)NLO contributions obtained in
this approach. In Section~\ref{sec:res} we first critically assess the
quality of the approximate results and discuss how we assign a
reasonable theory error to our results. We then present our new
approximate NNLO results for several phenomenologically interesting
observables for the LHC (at 8 and 14 TeV) and the Tevatron for a
typical cut-based analysis. Finally, in Section~\ref{sec:conclusions}
we briefly discuss some realistic future improvements that can be made
to our approach and end with some concluding remarks.

\section{Theoretical setup }
\label{sec:calc}

Initially, the standard technique to study RG-improved predictions for
hard-scattering processes at hadron colliders has been to work in
Mellin moment space. However, over the last few years an alternative
method \cite{Becher:2006nr,Becher:2006mr} based on SCET was developed.
This formalism allows to resum, directly in momentum space, soft gluon
emissions to all orders in the strong coupling constant $\alpha_s$.
This method was successfully applied to several collider processes,
and, in particular, to the pair production of heavy quarks at the
Tevatron and LHC~\cite{Ahrens:2010zv,Ahrens:2011mw}.  The formalism is
rather flexible since it allows to obtain resummed predictions at
the level of the differential distributions.

The SCET approach is based on the factorization of the partonic cross
section in the soft limit, in which it can be written as a convolution
of hard and soft functions. The hard functions contain the effects of
the virtual corrections, while the soft functions describe the
emission of the soft real radiation in the scattering processes.  In
fixed-order perturbation theory, partonic cross sections contain both
regular terms and singular distributions. The singular pieces can be
expressed as functions of a soft parameter which is closely related to
the energy of the soft gluons emitted. The details of the soft limit
are sensitive to the particular kinematic setup considered.  In
\cite{Ahrens:2010zv} the PIM kinematics was employed to investigate
the invariant mass distribution of the top pair, while in
\cite{Ahrens:2011mw} the 1PI kinematics was used to derive improved
predictions for transverse momentum and rapidity distributions of the
(anti-)top quark.

In \cite{Ahrens:2010zv} it was shown that after the convolution of the
hard-scattering kernels with the parton luminosities, the singular
terms in the threshold region generally contribute for more than
$90\%$ of the total NLO $t\bar{t}$ cross section.  In the threshold
region the contributions of the singular terms become large and
therefore they eventually need to be resummed to all orders in
perturbation theory. The enhancement of the singular terms in the
cross section is known to be due to the steeply falling behaviour of
the parton luminosities outside the threshold region. This effect is
called dynamical threshold enhancement.

In this section we will summarize the main results of
\cite{Ahrens:2010zv,Ahrens:2011mw} which are important for our work
and we will show how to extend these results to include the
semi-leptonic decays of the top quarks.  In
\cite{Ahrens:2010zv,Ahrens:2011mw} the authors studied the process
\begin{align}\label{eq:ttnodec}
N_1(P_1) + N_2(P_2)\,\to&\,t(p_t) + \bar{t}(p_{\bar{t}}) + X\, ,
\end{align}
where $N_1$ and $N_2$ are the incoming protons in the case of the LHC,
or proton anti-proton in the case of the Tevatron and $X$ is an
inclusive hadronic final state. In these works the top quarks were
always treated as stable on-shell partons and their decay was
omitted. 

The NWA, where the heavy particles produced in the scattering
process are allowed to decay to the physical final states whilst remaining
on-shell, is the first natural step to go beyond the stable top approximation.
In this limit the amplitudes describing the process factorize
into a part describing the production of a pair of on-shell top quarks
and parts describing the decay of an on-shell top and anti-top quark.  
In the NWA the radiative corrections to the process are given
by the factorizable corrections to the production and decay. In this
work we use the SCET approach to perturbatively improve the production 
subprocess beyond NLO, while keeping the accuracy of the decay 
subprocess at fixed NLO.

At LO in perturbation theory the two partonic channels 
contributing to the production subprocess in Eq.~(\ref{eq:ttnodec}) are
\begin{align}
q(p_1) + \bar{q}(p_2) \,\rightarrow\, t(p_t) + \bar{t}(p_{\bar{t}})\, ,\\
g(p_1) + g(p_2) \,\rightarrow\, t(p_t) + \bar{t}(p_{\bar{t}})\, ,
\end{align}
where the momenta of the incoming partons $p_i$($i=1,2$) are related
to the momenta of the initial state hadrons via the relation $p_i=x_i
P_i$.  We also introduce some kinematical invariants which are 
important for the description of the partonic scattering subprocesses:
\begin{align}
& s =  (p_1 + p_2)^2, \quad t_1 = (p_1-p_t)^2 - m^2_t, \quad u_1 =  (p_2-p_t)^2 - m^2_t\, ,\nonumber \\
& \quad \quad M^2 = (p_t + p_{\bar{t}})^2,\quad s_4 = s + t_1 + u_1 = (p_{\bar{t}}+k)^2-m^2_t\, ,
\end{align}
where $k$ is the 4-momentum of the hard-scattering products
different from the $t\bar{t}$ pair.
At Born level momentum conservation implies that $M^2=s$ and $s_4=0$.

\subsection{PIM and 1PI kinematics}
\label{subsec:pim1pi}

In this subsection we briefly summarize the main features of the PIM and 1PI
kinematics setups.
In the case of PIM kinematics it is convenient to introduce the quantities
\begin{align}
z=\frac{M^2}{s}\, ,\quad \quad \beta_t = \sqrt{1-\frac{4 m^2_t}{M^2}}\, ,
\end{align}
where the threshold region is identified by the limit $z \to 1$.  The
doubly differential partonic cross section in $M$ and in $\theta$ (the
scattering angle of the top quark in the partonic centre of mass
frame) takes the following form
\begin{align}\label{eq:partcs-pim}
\frac{d^2 \hat{\sigma}}{dM d\cos\theta}=
\frac{\pi \beta_t}{s M}\, \sum_{i,j}C_{{\rm{PIM}},\,ij}(z,M,m_t,\cos\theta,\mu_f)\, ,
\end{align}
where $\mu_f$ is the factorization scale and the sum is over all
possible incoming partons ($i,j$). The functions $C_{{\rm{PIM}},\,ij}$ in
Eq.~(\ref{eq:partcs-pim}) are usually called hard-scattering kernels and
they can be computed perturbatively. In fact, independently of the
kinematics, the kernels $C_{ij}$ have an expansion in powers of
$\alpha_s$
\begin{align}\label{eq:kernels}
C_{ij} = \alpha^2_s\left[C^{(0)}_{ij}+
\frac{\alpha_s}{4 \pi} C^{(1)}_{ij} +  
\left(\frac{\alpha_s}{4 \pi}\right)^2 C^{(2)}_{ij} + 
\mathcal{O}(\alpha^3_s)\right]\, .
\end{align}
At LO in $\alpha_s$, the non-zero scattering kernels correspond to the
quark anti-quark annihilation channel and to the gluon fusion channel,
$i,j \in \{q\bar{q}, gg\}$.  At higher orders in $\alpha_s$ one has to
consider the virtual and real corrections to the Born approximation
as well as the new partonic channels that open up at that order, for example
$g q\to t \bar{t} q$ at order $\alpha_s$.  The hard-gluon emission and
the additional production channel contribute to the NLO part of our
calculation and are taken into account exactly via the matching
procedure. Formally, near the partonic threshold, these contributions 
are suppressed by powers of the soft expansion parameter and 
therefore can be neglected in this region.

In \cite{Ahrens:2010zv} it was shown that in the limit of soft gluon emissions,
$z\to 1$, the hard-scattering kernels $C_{ij}$ factorize into a product
of hard and soft functions which are matrices in colour space
\begin{align}
C_{{\rm{PIM}},\,ij}(z,M,m_t,\cos \theta,\mu_f) = 
{\rm Tr} \left[ {\bf H}_{ij}(M,m_t,\cos\theta,\mu_f) \, 
{\bf S}_{{\rm{PIM}},\,ij}(\sqrt{s}(1-z),M,m_t,\cos\theta,\mu_f)\right] \, .
\end{align}
The hard functions ${\bf H}_{ij}$ are obtained from the virtual
corrections and they are ordinary functions of their arguments. The
soft functions ${\bf S}_{{\rm{PIM}},\,ij}$ describe the real emission
of soft gluons and contain distributions which become singular in the
threshold limit $z\to 1$. At order $\alpha^n_s$ the soft functions
depend on terms proportional to plus-distributions of the form
\begin{align} \label{eq:pim-plus-dist}
P_m(z) = \left[\frac{\ln^m(1-z)}{1-z}\right]_+\,;\quad m=0,\ldots,2 n-1\, ,
\end{align}
and on terms proportional to the delta distribution $\delta(1-z)$.

The 1PI kinematics is commonly used to describe observables related to
a single heavy particle rather than a pair. At the partonic level, 1PI
kinematics allows one to write the top quark rapidity $(y)$ and transverse
momentum $(p_{T})$ distributions as
\begin{align}\label{eq:partcs-1pi}
\frac{d^2 \hat{\sigma}}{dp_T dy}= \frac{2 \pi p_T}{s} \, 
\sum_{i,j}C_{{\rm{1PI}},\,ij}(s_4,s,t_1,u_1,m_t,\mu_f)\, .
\end{align}
As in the PIM case, in 1PI kinematics the hard-scattering kernels factorize 
into a product of hard and soft functions in the limit of soft gluon 
emissions $s_4\to 0$
\begin{align}
C_{{\rm{1PI}},\,ij}(s_4,s,t_1,u_1,m_t,\mu_f) = 
{\rm Tr} \left[ {\bf H}_{ij}(s,t_1,u_1,m_t,\mu_f) \, 
{\bf S}_{{\rm{1PI}},\,ij}(s_4,s,t_1,u_1,m_t,\mu_f)\right]\, ,
\end{align}
where again ${\bf H}_{ij}$ and ${\bf S}_{{\rm{1PI}},\,ij}$ are matrices in
colour space.  Whilst the hard functions in PIM and 1PI kinematics are
identical, the soft functions are different.  In
particular, at order $\alpha^n_s$ the 1PI soft functions include terms
proportional to singular plus-distributions which depend on $s_4$
\begin{align} \label{eq:1pi-plus-dist}
\bar{P}_m(s_4) = \left[\frac{\ln^m(s_4/m^2_t)}{s_4}\right]_+ = 
\frac{1}{m^2_t} P_m\left(1-\frac{s_4}{m^2_t}\right)\,;\quad m=0,\ldots,2 n-1\, .
\end{align}

\subsection{Inclusion of tree-level decays}

Here we describe how we extend the formalism outlined in the previous 
subsection in order to include the tree-level semi-leptonic decays of 
the top quarks
\begin{align}
t \,\to\, b W^+\,\to\, b(p_3)\bar{l}_1(p_5)\nu_1(p_6)\, ,\nonumber\\
\bar{t}\, \to\, \bar{b} W^- \, \to\, \bar{b}(p_4)l_2(p_7)\bar{\nu}_2(p_8)\, ,
\end{align}
where
\begin{align}
\label{eq:topmom}
p_t = p_3 + p_5 + p_6,\quad & p_{\bar{t}} = p_4 + p_7 + p_8 \, ,\\
\label{eq:wmom}
p_{W^+} = p_5+p_6, \quad & p_{W^-} = p_7+p_8\, .
\end{align}

Given that we are interested in keeping full spin-correlation effects,
the spins of the top and anti-top quarks must not be summed
over. Unfortunately the results available in the literature for the
$t\bar{t}$ hard functions, ${\bf H}_{ij}$, are already summed over the
spins of the external particles and do not allow for a direct inclusion
of the spin-correlated top decays. For this reason the hard functions
had to be computed anew keeping the spin information explicit.  This
was achieved using spinor-helicity methods, and in particular making
use of the one-loop helicity amplitudes for the $t\bar{t}$ production
process which were recently computed in \cite{Badger:2011yu}.  Using
the standard procedure for sewing together helicity amplitudes for the
production of a heavy quark pair with the appropriate helicity
amplitudes for the decay subprocess (and summing over the helicities
of the heavy quarks), it was possible to construct the required hard
functions where the spin-correlations are included. For each helicity
configuration of the eight external particles,
$\{\lambda\}\equiv(\lambda_1,\ldots,\lambda_8)$, the amplitude can be
expressed as a colour-decomposed function of eight external momenta
$\{p_i \}_1^8\equiv (p_1,\dots, p_8)$
\begin{align}\label{eq:ampcolordec}
\mathcal{M}^{\{\lambda\}}_{ij,\, \{a\}} (\{p_i \}_1^8,m_t,\mu_f) = 
\sum_I \mathcal{M}^{\{\lambda\}}_{ij,\, I} (\{p_i \}_1^8,m_t,\mu_f) 
(c^{ij}_I)_{\{a\}}\, .
\end{align}
The $(c^{ij}_I)_{\{a\}}$ are independent colour structures which
represent orthogonal basis vectors in the space of colour-singlet
amplitudes and $\{a\} \equiv (a_1,a_2,a_3,a_4)$ is the set of colour
indices relative to the external coloured particles which can be in
the fundamental or adjoint representation. In the NWA,
when the decay is added at tree-level, the colour indices carried by
the $b$-quarks are equal to the colour indices of the top quarks to
which they are attached, therefore $a_3 = a_t$ and $a_4=a_{\bar{t}}$.
Explicitly, the colour basis is chosen to be
\begin{align}
(c^{q\bar{q}}_1)_{\{a\}} = \delta_{a_1 a_2}\delta_{a_3 a_4}, 
&\quad \quad (c^{q\bar{q}}_2)_{\{a\}} = t^c_{a_2 a_1} t^c_{a_3 a_4}\, ,\nonumber\\
(c^{gg}_1)_{\{a\}} = \delta^{a_1 a_2}\delta_{a_3 a_4}, 
\quad\quad (c^{gg}_2)_{\{a\}} &= i f^{a_1 a_2 c} \, t^c_{a_3 a_4},
\quad\quad (c^{gg}_3)_{\{a\}} = d^{a_1 a_2 c}\, t^c_{a_3 a_4}\, ,
\end{align}
where the SU($N$) matrices $t^a$ satisfy the usual relations
$[t^a,t^b]=i f^{abc}\,  t^c$ and $\{t^a,t^b\}=d^{abc}\, t^c +
\delta^{ab}/N$. In order to construct the hard function matrices the
infrared (IR) poles must be removed from the QCD amplitudes. This is
done following the procedure in \cite{Becher:2009qa} which, in short,
can be written as
\begin{align}\label{eq:renamp}
\mathcal{M}_{ij,\,\{a\}}^{\text{ren},\,\{\lambda\}} \equiv 
\lim_{\epsilon \to 0} {\bf Z }^{-1}_{ij}(\epsilon)
\mathcal{M}^{\{\lambda\}}_{ij,\,\{a\}}(\epsilon) = 
4\pi\alpha_s \left[ \mathcal{M}_{ij,\,\{a\}}^{\text{ren}\,(0),\,\{\lambda\}} +
\frac{\alpha_s}{4\pi}\mathcal{M}_{ij,\,\{a\}}^{\text{ren}\,(1),\,\{\lambda\}} + 
\dots \right] \, ,
\end{align}
where the IR poles are removed by the matrices ${\bf Z
}^{-1}_{ij}(\epsilon)$ \cite{Becher:2009kw,Ferroglia:2009ii}. 

The hard function matrices have a perturbative expansion in $\alpha_s$
\begin{align}
{\bf H}_{ij} = \alpha^2_s \frac{1}{d_R} \left( {\bf H}^{(0)}_{ij} + 
\frac{\alpha_s}{4 \pi} {\bf H}^{(1)}_{ij} + \ldots \right)\, ,
\end{align}
where $d_R = N$ for the $q\bar{q}$ channel and $d_R = N^2-1$ for the
$gg$ channel.  The matrix elements $H_{IJ}$ (for simplicity we drop
the subscript $ij$ labelling the production channel) can be expressed
in terms of the renormalized colour-decomposed QCD amplitudes as
\begin{align}
H^{(0)}_{IJ} &= \frac{1}{4} \sum_{\{\lambda\}} 
\left(\mathcal{M}^{\text{ren}\,(0)\, \{\lambda\}}_I\right)^* 
\left(\mathcal{M}^{\text{ren}\,(0)\, \{\lambda\}}_J\right)\, ,\\
H^{(1)}_{IJ} &= \frac{1}{4} \sum_{\{\lambda\}} \left[ 
\left(\mathcal{M}^{\text{ren}\,(0)\, \{\lambda\}}_I\right)^* 
\left(\mathcal{M}^{\text{ren}\,(1)\, \{\lambda\}}_J\right) + 
\left(\mathcal{M}^{\text{ren}\,(1)\, \{\lambda\}}_I\right)^* 
\left(\mathcal{M}^{\text{ren}\,(0)\, \{\lambda\}}_J\right)\right]  \, ,
\end{align}
where we have summed over all possible external helicities. After
including the tree-level decays, the hard functions ${\bf
  H}_{ij}(\{p_i \}_1^8,m_t,\mu_f)$ depend explicitly on the external
momenta which are subject to the constraints in
\Eqns{eq:topmom}{eq:wmom}.  It would be straightforward to extend
these results and compute the hard functions where the one-loop
corrections to the decay are included in the NWA.  This would be
needed to implement some improved prediction at the decay level, but
for the moment we focus on adding the decay corrections strictly at
fixed order (LO/NLO). Hence the computation of these additional pieces
is not needed for the present work.

Since we work in the NWA and the decay is added at fixed order, the
soft functions ${\bf S}_{{\rm{PIM}}}$ and ${\bf S}_{{\rm{1PI}}}$ are
not changed by the inclusion of the tree-level top decays. Therefore,
we make use of the results for the PIM and 1PI NLO soft functions
which are available in the literature. The calculation of the PIM soft
functions at NLO was carried out in \cite{Ahrens:2010zv} while the
results for the 1PI soft functions can be found in
\cite{Ahrens:2011mw}.

\subsection{Structure of the hard-scattering kernels}
\label{subsec:structurehsk}

The hard ${\bf{H}}_{ij}$ and soft ${\bf{S}}_{ij}$ functions for the
production subprocess satisfy renormalization-group equations (RGEs)
whose precise form was derived in \cite{Ahrens:2010zv} for the PIM
case and in \cite{Ahrens:2011mw} for 1PI. The relevant two- and
three-loop anomalous dimensions which enter the RGEs can be found in
\cite{Ferroglia:2009ii,Korchemskaya:1992je,Moch:2004pa}. Amongst
these, the two-loop anomalous dimensions for massive partons
\cite{Ferroglia:2009ii} are a key ingredient of this approach. Given
full knowledge of the NLO hard and soft functions, it is possible to
exploit the information from the RGEs to derive fixed-order
approximate $\mathcal{O}({\alpha^2_s})$ expressions for the
hard-scattering kernels. The procedure to obtain approximate NNLO
results from the NNLL resummation formula was described in detail in
\cite{Ahrens:2010zv,Ahrens:2011mw}. We do not repeat the derivation
here and instead focus on the results of this procedure that are
relevant for our case.  In general the NNLO hard-scattering kernels
have the following expansion in terms of hard and soft functions
\begin{align}\label{eq:hsk2}
C^{(2)}_{ij} = 
{\rm{Tr}}\left[{\bf {H}}^{(1)}_{ij}{\bf {S}}^{(1)}_{ij}\right] + 
{\rm{Tr}}\left[{\bf {H}}^{(0)}_{ij}{\bf {S}}^{(2)}_{ij}\right] + 
{\rm{Tr}}\left[{\bf {H}}^{(2)}_{ij}{\bf {S}}^{(0)}_{ij}\right]\, ,
\end{align}
where currently only the first term on the r.h.s. of \Eqn{eq:hsk2} is
known exactly. As for the second term, only the logarithmic parts of
${\bf {S}}^{(2)}_{ij}$ are known completely and can be obtained by
solving the RGE for the soft functions at $\mathcal{O}({\alpha^2_s})$.
In principle, the scale-dependent parts of ${\bf {H}}^{(2)}_{ij}$,
relevant to the third term in the sum, can also be computed via
RGE. However, the inclusion of these terms is problematic as this
cannot be done in an unambiguous way and might introduce an artificial
reduction of the scale dependence. Hence, we follow the choice made in
\cite{Ahrens:2011mw} to drop these terms altogether. 

In momentum space the NNLO hard-scattering kernels in \Eqn{eq:kernels}, 
have the following structure in the case of PIM kinematics
\begin{align}
\label{eq:pimkernels}
C^{(2)}_{\rm{PIM}}(z,\{p_i \}_1^8,m_t,\mu_f) &=  
\sum^{3}_{m=0} D^{(2)}_{{\rm{PIM}},\,m}(z,\{p_i \}_1^8,m_t,\mu_f) P_m(z) \nonumber\\
&+ Q^{(2)}_{\rm{PIM},\,0}(\{p_i \}_1^8,m_t,\mu_f)\delta(1-z) +
R^{(2)}_{\rm{PIM}}(z,\{p_i \}_1^8,m_t,\mu_f)    \, ,
\end{align}
while in the case of 1PI kinematics we have
\begin{align}
\label{eq:1pikernels}
C^{(2)}_{\rm{1PI}}(s_4,\{p_i \}_1^8,m_t,\mu_f) &=  
\sum^{3}_{m=0} D^{(2)}_{{\rm{1PI}},\,m}(s_4,\{p_i \}_1^8,m_t,\mu_f) \bar{P}_m(s_4) \nonumber\\
&+ Q^{(2)}_{\rm{1PI},\,0}(\{p_i \}_1^8,m_t,\mu_f)\delta(s_4) + 
R^{(2)}_{\rm{1PI}}(s_4,\{p_i \}_1^8,m_t,\mu_f)    \, .
\end{align}
The expressions of the coefficients multiplying the plus-distribution
to $\mathcal{O}({\alpha^2_s})$, $D^{(2)}_{{\rm{PIM}},\,m}$ and
$D^{(2)}_{{\rm{1PI}},\,m}$ where $m=0,\ldots,3$, are known exactly.
They depend on the momenta of the external particles $\{p_i \}_1^8$,
on the variables $z$ and $s_4$ for PIM and 1PI respectively, on the
top mass $m_t$ and on the scale $\mu_f$.  The coefficients of the
$\delta$-function are only partly known, with the missing terms coming
from the constant pieces of the two-loop soft functions and the
scale-independent parts of the two-loop virtual corrections. The
former pieces are currently unknown while the situation has recently
changed for the latter. Namely, the two-loop amplitudes for stable top
production have become available in the form of a large numerical grid
\cite{Baernreuther:2013caa}.  However, the results are not given in a
helicity-amplitude-structure-like form, therefore are not (yet)
suitable for a direct inclusion of the top decays.  The functions
$R^{(2)}_{\rm{PIM}}(z,m_t,\mu_f)$ and
$R^{(2)}_{\rm{1PI}}(s_4,m_t,\mu_f)$ are finite in the limit $z\to 1$
and $s_4\to 0$ respectively. These remainders do not include all the
regular terms at $\mathcal{O}({\alpha^4_s})$ since the total regular
contributions could only be obtained with a full calculation at NNLO
accuracy.

Finally the hard-scattering kernels in
\Eqns{eq:pimkernels}{eq:1pikernels} can be integrated over the
appropriate phase space to get the partonic cross sections
contributions in PIM and 1PI kinematics at order
$\mathcal{O}(\alpha^4_s)$
\begin{align}
\hat{\sigma}^{(2)}_{{\rm{PIM}}} &=\frac{1}{2 s} \int d \Phi_{t\bar{t},\,\rm{PIM}} \,
 C^{(2)}_{\rm{PIM}}(z,\{p_i \}_1^8,m_t,\mu_f) \,, \\
\hat{\sigma}^{(2)}_{{\rm{1PI}}} &= \frac{1}{2 s} \int d \Phi_{t\bar{t},\,\rm{1PI}} \, 
C^{(2)}_{\rm{1PI}}(s_4,\{p_i \}_1^8,m_t,\mu_f) \, .
\end{align}
These expressions are in a form that is suitable for Monte Carlo
integration.

\subsection{Monte Carlo Implementation}
\label{subsec:mc-implementation}

In this subsection we briefly summarize the MC
integration method and give details on the implementation of the
approximate (N)NLO formulas in the MC framework.  The starting point for
a MC integration of a final-state phase space is the formula for the
cross section
\begin{align} \label{eq:basic-sigma-formula}
\sigma &= \int\, dx_1 dx_2 \,
f_1(x_1,\mu_f)f_2(x_2,\mu_f) \int d\Phi_n[s] 
\frac{d\hat{\sigma}}{d\Phi_n}(x_1,x_2) \nonumber \\
& = \frac{1}{2 S} \int \, dx_1 dx_2 
\frac{f_1(x_1,\mu_f)f_2(x_2,\mu_f)}{x_1 x_2} 
\int d\Phi_n[s]\, |\mathcal{M}(\{p_i \}^n_1)|^2 \; 
\mathcal{F}(\{p_i\}^n_1),
\end{align}
where $d\Phi_n$ is the usual final-state phase-space measure for $n$
final state particles,
\begin{align}\label{eq:standard-phase-space}
d\Phi_n[s] = 
\left(\prod^n_{i=1} \frac{d^3 p_i}{(2\pi)^3 \, 2E_i} \,  \right) 
\delta^{4}( p_1+p_2 - p_1 - \ldots - p_n ),    
\end{align}
$|\mathcal{M}|^2$ is the squared matrix element for the process considered,
$f_{1,2}$ are the parton distribution functions (PDFs), $S$ is the
hadronic centre-of-mass energy squared and $\mathcal{F}$ is the
measurement function, which defines the (infrared (IR) safe)
observable to be studied.  The motivation to perform the above
phase-space integrals numerically comes when these become
prohibitively difficult to compute analytically.  This can happen both
when the final state consists of many particles and when one is
interested in imposing conditions on the final state (i.e. when
$\mathcal{F}$ is different from the identity, for example, to account
for finite reaches of detectors, reconstructing jets out of
final-state partons etc.).

The most convenient way of performing such integrals numerically is via
MC integration. Random numbers are generated and mapped to phase-space
points $\{p_i \}^n_1$ which are then used to evaluate the
matrix-element. This procedure yields a weight which can then be
binned into histograms according to any distribution one wishes to
study. The sum over all weights over the full range of possible bins
of each observable is equal to the total cross section. It is clear
that supplementing the phase space, \Eqn{eq:standard-phase-space},
with conditions on the final state does not complicate the MC
integration, in the sense that to achieve this it is simply necessary
to throw away (not bin) the weights whose associated phase-space
configuration does not satisfy the given
conditions.\footnote{Implementations of such MC integrations are also
  usually termed `parton-level' Monte-Carlo integrators since no effects
  showering of external partons and hadronization are included.}

Beyond LO the numerical phase-space integration becomes a little more
difficult to deal with due to the different phase spaces the virtual and
real corrections live in (at NLO these are $d\Phi_n$ and $d\Phi_{n+1}$
respectively). The integrations over these phase spaces are
individually divergent, but it is well-known that divergences cancel
in the sum. For NLO there exist standard methods that achieve this
cancellation numerically and in our implementation of NLO corrections
we have employed both the FKS \cite{Frixione:1995ms,Frederix:2009yq}
and Catani-Seymour \cite{Catani:1996vz,Catani:2002hc} subtraction
methods (the complete agreement of the results of the two methods is a
strong check of our implementation).

The hard-scattering kernels described in
Sections~\ref{subsec:pim1pi}--\ref{subsec:structurehsk} can be used in
\Eqn{eq:basic-sigma-formula} in a similar way as the standard LO or
NLO hard-scattering matrix elements. In a MC-integration approach the
only change with respect to the kernels of \cite{Ahrens:2010zv,
  Ahrens:2011mw} being that we have now expressed them explicitly as
functions of the outgoing particle momenta.  We obtain approximate
(N)NLO results using these hard-scattering kernels derived from both
PIM and 1PI kinematics, which, apart from the kernels themselves,
differ in precisely what approximations are made in the phase-space
integration, $d\Phi_{n,\text{PIM}}$ or $d\Phi_{n,\text{1PI}}$.

\subsubsection{Monte Carlo and PIM}

The PIM cross-section formula is
\begin{align}\label{eq:pim-sigma}
\sigma =& \frac{1}{2 S}\int d\hat\Phi_{t\bar{t}} \, 
\int^1_\tau \frac{dz}{z} \, \int^1_{\tau/z} \frac{dx}{x}\, 
f_1(x, \mu_f) f_2(\tau/(x\,z), \mu_f)\, C_{\rm{PIM}}(z,\{p_i(x,z)\}_1^8, m_t, \mu_f),
\end{align}
where $\tau = (p_t + p_{\bar{t}})^2/S$ and we have replaced the
integration on $M$ and $\cos\theta$ in \Eqn{eq:partcs-pim} with the
full phase space of the $t\bar{t}$ decay products, $d\hat\Phi_{t\bar{t}}$. To
make a connection between \Eqn{eq:pim-sigma} and the general
expression given in \Eqn{eq:basic-sigma-formula} we note that
$d\hat\Phi_{t\bar{t}}$ can be obtained by starting from $d\Phi_{n=3}$
for a $t,\bar{t},g$ final state and integrating over the angular
variables of the (soft) gluon. If the decay of the (anti-)top is to be
included, $d\hat\Phi_{t\bar{t}}$ also contains the corresponding
angular integrations. For simplicity we think of the hard-scattering
kernels $C_{\rm{PIM}}$ as being multiplied by the measurement function
$\mathcal{F}$ introduced in \Eqn{eq:basic-sigma-formula}. It is
important to note that in order to exactly reproduce the phase-space
integration found in \cite{Ahrens:2010zv} (i.e. change the phase space
from the `full' to the `approximate' one, $d\hat\Phi_{t\bar{t}} \to
d\hat\Phi_{t\bar{t},\,\text{PIM}}$) an additional factor of $1/z$ must
be included in the $z$-integral. This comes about in the derivation of
the factorization formula when two separate factors of $z^{1/2}$ have
been set equal to 1, which, of course, is perfectly acceptable in the
limit $z\to 1$, in which the factorization formula is strictly true.
When performing the integral using an MC method, introducing this
additional factor of $1/z$ allows us to numerically reproduce the
differential results in \cite{Ahrens:2010zv}.

Using the change of variables $x_1 = x$, $x_2 = \tau/(x\,z)$ and $ v =
z$ it is possible to rewrite the cross section in the form
\begin{align} \label{eq:pim-implem-eq}
\sigma &= \frac{1}{2 S} \int \frac{dx_1}{x_1} 
\frac{dx_2}{x_2} f_1(x_1,\mu_f) f_2(x_2,\mu_f) \int^{1}_{\frac{4 m^2_t}{x_1 x_2 S}} dv\,
\int d{\Phi}_{t\bar{t}}[v s] \, 
C_{\rm{PIM}}(v,\{ p_i(x_1,x_2,v)\}_1^8, m_t, \mu_f),  
\end{align}
where we have explicitly extracted the integration variable $v$ from
the complete $d\hat\Phi_{t\bar{t},\,\text{PIM}}$ phase space and the
remaining integrations $d\Phi_{t\bar{t}}[v s]$ now depend on the
reduced partonic centre of mass energy squared $v s$.  This way of
writing the phase-space integration is perhaps more familiar for use
in MCs, since this is the typical form of the standard NLO
initial-state collinear contributions. We have made it clear above
that the hard-scattering kernel is now evaluated using momentum
configurations that depend on the variable $v$, which is the argument
of the PIM plus-distributions of \Eqn{eq:pim-plus-dist}.

\subsubsection{Monte Carlo and 1PI}

For 1PI, the formula we have implemented has a somewhat different
structure. The starting point for our MC implementation are Eqs.~(42)
and (43) of \cite{Ahrens:2011mw}, which can be formulated in the
following form
\begin{align} \label{eq:starting-1pi-eq}
\sigma &= \frac{1}{4 S} \int d\hat\Phi_{t\bar{t},\,\text{1PI}} \nonumber \\ 
&\left( \int \frac{dx_1}{x_1}  \int^{s^{\text{max},t}_4}_0 d s_4 \,
\frac{f_1(x_1,\mu_f) f_2(x'_2(s_4),\mu_f)}{s_4- x_1 t_1}
\, C_{\rm{1PI}}\left(s_4,\{ p_i(x_1,x'_2(s_4)\}_1^8,m_t,\mu_f)\right) \right. \nonumber \\  
+ &\left. \int \frac{dx_2}{x_2}  \int^{s^{\text{max},u}_4}_0 d s_4 \, 
\frac{f_1(x'_1(s_4),\mu_f) f_2(x_2,\mu_f)}{s_4- x_2 u_1}
\, C_{\rm{1PI}}(s_4,\{ p_i(x'_1(s_4),x_2\}_1^8,m_t,\mu_f)) \right),
\end{align}
once the dependence on $p_T(t)$ and $y(t)$ is replaced by the
phase space of the $t\bar{t}$ decay products, $d\hat\Phi_{t\bar{t},\,\rm{1PI}}$.
To obtain \Eqn{eq:starting-1pi-eq}, a change of variables is performed by
trading $x_1$ or $x_2$ in favour of $s_4$. To do so we introduce the functions
\begin{align}
x'_1(s_4) &\equiv \frac{s_4 - x_2 U_1}{x_2 S + T_1} \;\; 
\text{and} \;\; x'_2(s_4) \equiv \frac{s_4 - x_1 T_1}{x_1 S + U_1}
\end{align}
and a set of hadronic kinematical variables related to the partonic ones
\begin{align}
S=\frac{s}{x_1 x_2,}\quad \quad T_1 = \frac{t_1}{x_1}, \quad \quad U_1 = \frac{u_1}{x_2}\, .
\end{align}
The upper bound on the $s_4$ integration is found to be
\begin{align}
s^{\text{max},t} &= x_1(S+ T_1) + U_1,  \;\;\;\; s^{\text{max},u} = x_2( S + U_1) + T_1.
\end{align}
The cross section in \Eqn{eq:starting-1pi-eq} is written as a
sum of two terms (related via the exchange $t_1 \leftrightarrow u_1$)
in order to preserve the symmetry of the gluon channel under the
interchange $y(t) \to -y(t)$.  Before performing the integrals using
the MC method, it is important (in order to match our results with
those of \cite{Ahrens:2011mw}), to make sure the same approximations
are made in all relevant places. In \cite{Ahrens:2011mw} the choice
was made to keep \emph{only} the explicit dependence on $s_4$ non-zero
in the first argument of the $C_{\text{1PI}}$-functions.  This means
that momenta (phase space) we generate and use to evaluate the
hard-scattering kernels must be of `Born-level' kinematics, i.e. we
generate momenta satisfying
\begin{align}
\{ p_i(x_1,x'_2(s_4=0))\}_1^8\,, \quad\quad \{ p_i(x'_1(s_4=0),x_2\}_1^8\, ,
\end{align}
respectively for the first and second term in the sum in
\Eqn{eq:starting-1pi-eq} and we evaluate the hard-scattering
kernels with these momenta as
\begin{align}
C_{\text{1PI}}\left(s_4,\{ p_i(x_1,x'_2(s_4=0))\}_1^8,m_t,\mu_f\right)
\,, \; 
& \;\; C_{\text{1PI}}\left(s_4,\{ p_i(x'_1(s_4=0),x_2)\}_1^8, m_t, \mu_f\right).
\end{align}  
Note though, that where there is explicit dependence on the variable
$s_4$ (for example, in the plus-distributions ,
\Eqn{eq:1pi-plus-dist}), this must of course be evaluated keeping $s_4
\neq 0$.  Making the further choice of exchanging one of the
kinematical variables inside $d\hat\Phi_{t\bar{t},\,\text{1PI}}$ with
the PDF momentum fraction $x_2$ makes it possible to explicitly
extract this integration variable and rewrite the cross section in
terms of the remaining phase-space integrations $d\Phi_{t\bar{t}}$
\begin{align} \label{eq:1pi-implem-eq}
\sigma &= \frac{1}{4S}  \int dx_1 dx_2 
\int d\Phi_{t\bar{t}}[s]  \nonumber \\ 
&\left( \int^{s^{\text{max},t}_4}_0 d s_4
\frac{f_1(x_1,\mu_f) f_2(x'_2(s_4),\mu_f)}{x_1 x'_2(s_4)}
C_{\text{1PI}}\left(s_4,\{ p_i(x_1,x'_2(0))\}_1^8,m_t,\mu_f\right) \right. \nonumber \\  
+ &\left. \int^{s^{\text{max},u}_4}_0 d s_4 
\frac{f_1(x'_1(s_4),\mu_f) f_2(x_2,\mu_f)}{x'_1(s_4) x_2}
C_{\text{1PI}}\left(s_4,\{ p_i(x'_1(0),x_2)\}_1^8, m_t, \mu_f\right)\right)\, ,
\end{align}
where the upper bounds on the $s_4$ integrations have to be expressed
in terms of the new integration variables.  Evaluating the 1PI cross
section in \Eqn{eq:1pi-implem-eq} using MC methods, we find complete
agreement with the distributions of \cite{Ahrens:2011mw}.  It must be
emphasised that there is significant freedom in the final form chosen
to implement. As has been discussed in depth in \cite{Ahrens:2011mw},
this depends heavily on where the $s_4 = 0$ approximation is
enforced. In the MC approach this feeds into how the phase space is
generated and thus how the hard-scattering kernels are evaluated. The
various choices of course all differ by formally subleading,
$\mathcal{O}(s_4)$, contributions which nonetheless can have
significant numerical impact. For this work we have made the choice to
treat the subleading corrections in a way such that we recover the
differential results of \cite{Ahrens:2011mw}.

\subsubsection{Flexibility of Monte Carlo}

It should be clear that \Eqns{eq:pim-implem-eq}{eq:1pi-implem-eq}
enable us to obtain differential cross sections for top-quark
pair production at approximate (N)NLO accuracy by using
$C^{(1)}_{\text{PIM/1PI}}$ ($C^{(2)}_{\text{PIM/1PI}}$).  Since we
also wish to study the decay products of the top, the top-pair
phase space is straightforwardly extended to the phase space of
$W^+W^-b\bar{b}$-production (in the NWA for the top quarks) for these
approximate (N)NLO cross sections.

As has already been mentioned, the MC implementation naturally lends
itself to imposing conditions on the final state, clustering
final-state partons into jets etc. Thus, in this more flexible
framework it is possible to study the predictions of PIM and 1PI under
more realistic analysis setups. Strictly speaking, the PIM and 1PI
results should only be used to make predictions for
$\{M_{t\bar{t}},\cos\theta\}$ and $\{p_T(t),y(t)\}$ distributions,
since it is only for these that the analytic formulas are
formally valid.
However, given that the MC approach deals with weights
and momenta, it is possible to make a prediction for any distribution
using the PIM and 1PI kernels.  Of course, no factorization formula
has been derived for any of the `wrong' distributions we obtain or
indeed any of the `right' distributions, where analysis cuts have been
imposed on the final state. However, importantly, we have strong 
checks on when and where to trust the RG-improved results and we will 
argue that \emph{a posteriori} these new results are indeed sensible.

\subsubsection{Structure of results}

Given that we are working in the NWA, corrections to the production
and decay of the pair of on-shell top quarks can be treated
separately.  The (potential) improvement to the current
state-of-the-art NWA predictions \cite{Bernreuther:2004jv,
  Melnikov:2009dn, Campbell:2012uf} we explore here is the addition of
an approximate NNLO correction in the production subprocess. To aid
comprehension in the following section, we introduce the differential
cross sections $d\sigma^{xx}_{\text{prod}}$ and
$d\sigma^{xx}_{\text{full}}$. The former includes the decays of top
quarks at LO only, whilst the latter contains the NLO corrections to
the decays. Here, $xx$ will label the accuracy of the results,
i.e. this will be either the normal fixed-order LO or NLO results or
the approximate next-to-leading order (nLO) or approximate
next-to-next-to leading order (nNLO) results.  In detail, the sets of
differential cross sections that we will examine are
\begin{align}
d\sigma^{\text{nLO}}_{\text{prod}} &= 
(\Gamma^{\text{LO}}_t)^{-2} \left( d\sigma^{(0)}_{t\bar{t}} + 
d\tilde{\sigma}^{(1)}_{t\bar{t}} \right) 
\varotimes d\Gamma^{(0)}_{t\to l^+\nu_l b} \varotimes  
d\Gamma^{(0)}_{\bar{t}\to l^- \bar{\nu}_l \bar{b}} \\[5pt]
d\sigma^{\text{NLO}}_{\text{prod}} &= 
(\Gamma^{\text{LO}}_t)^{-2} \left( d\sigma^{(0)}_{t\bar{t}} + 
d\sigma^{(1)}_{t\bar{t}}  \right)
\varotimes d\Gamma^{(0)}_{t\to l^+\nu_l b} \varotimes  
d\Gamma^{(0)}_{\bar{t}\to l^- \bar{\nu}_l \bar{b}} \\[5pt]
d\sigma^{\text{nNLO}}_{\text{prod}} &= (\Gamma^{\text{LO}}_t)^{-2} 
\left( d\sigma^{(0)}_{t\bar{t}} + d\sigma^{(1)}_{t\bar{t}} +
d\tilde{\sigma}^{(2)}_{t\bar{t}} \right)
\varotimes d\Gamma^{(0)}_{t\to l^+\nu_l b} \varotimes  
d\Gamma^{(0)}_{\bar{t}\to l^- \bar{\nu}_l \bar{b}}
\end{align} 
and
\begin{align}
d\sigma^{\text{nLO}}_{\text{full}} = (\Gamma^{\text{NLO}}_t)^{-2} 
\left\{  \left( d\sigma^{(0)}_{t\bar{t}} + d\tilde{\sigma}^{(1)}_{t\bar{t}} \right) \right. 
& \varotimes d\Gamma^{(0)}_{t\to l^+\nu_l b} \varotimes  
d\Gamma^{(0)}_{\bar{t}\to l^- \bar{\nu}_l \bar{b}}  \nonumber \\[3pt]
 + d\sigma^{(0)}_{t\bar{t}} & \varotimes d\Gamma^{(1)}_{t\to l^+\nu_l b} \varotimes  
d\Gamma^{(0)}_{\bar{t}\to l^- \bar{\nu}_l \bar{b}} \nonumber \\[3pt]
 + d\sigma^{(0)}_{t\bar{t}} & \left. \varotimes \, d\Gamma^{(0)}_{t\to l^+\nu_l b} 
\varotimes  d\Gamma^{(1)}_{\bar{t}\to l^- \bar{\nu}_l \bar{b}}  \right\} \\[5pt] 
d\sigma^{\text{NLO}}_{\text{full}} = (\Gamma^{\text{NLO}}_t)^{-2} 
\left\{  \left( d\sigma^{(0)}_{t\bar{t}} + d\sigma^{(1)}_{t\bar{t}} \right) \right. 
& \varotimes d\Gamma^{(0)}_{t\to l^+\nu_l b} \varotimes  
d\Gamma^{(0)}_{\bar{t}\to l^- \bar{\nu}_l \bar{b}}  \nonumber \\[3pt]
 + d\sigma^{(0)}_{t\bar{t}} & \varotimes d\Gamma^{(1)}_{t\to l^+\nu_l b} 
\varotimes  d\Gamma^{(0)}_{\bar{t}\to l^- \bar{\nu}_l \bar{b}} \nonumber \\[3pt]
 + d\sigma^{(0)}_{t\bar{t}} & \left. \varotimes \, d\Gamma^{(0)}_{t\to l^+\nu_l b}
 \varotimes  d\Gamma^{(1)}_{\bar{t}\to l^- \bar{\nu}_l \bar{b}}  \right\} \\[5pt]
d\sigma^{\text{nNLO}}_{\text{full}} = (\Gamma^{\text{NLO}}_t)^{-2} 
\left\{  \left( d\sigma^{(0)}_{t\bar{t}} + d\sigma^{(1)}_{t\bar{t}} + 
d\tilde{\sigma}^{(2)}_{t\bar{t}}  \right) \right. 
& \varotimes d\Gamma^{(0)}_{t\to l^+\nu_l b} \varotimes  
d\Gamma^{(0)}_{\bar{t}\to l^- \bar{\nu}_l \bar{b}}  \nonumber \\[3pt]
 + d\sigma^{(0)}_{t\bar{t}} & \varotimes d\Gamma^{(1)}_{t\to l^+\nu_l b} 
\varotimes  d\Gamma^{(0)}_{\bar{t}\to l^- \bar{\nu}_l \bar{b}} \nonumber \\[3pt]
 + d\sigma^{(0)}_{t\bar{t}} & \left. \varotimes \, 
d\Gamma^{(0)}_{t\to l^+\nu_l b} \varotimes  d\Gamma^{(1)}_{\bar{t}\to l^- \bar{\nu}_l
   \bar{b}}  \right\}.
\label{eq:nNLO-full}
\end{align} 
Here, $d\sigma^{(0,1)}_{t\bar{t}}$, $d\Gamma^{(0,1)}_{t\to l^+\nu_l
  b}$ and $d\Gamma^{(0,1)}_{\bar{t}\to l^- \bar{\nu}_l \bar{b}}$ are
the standard LO and NLO differential production cross sections and
partial widths.  The approximate ${\cal O}(\alpha_s^3)$ NLO and ${\cal
  O}(\alpha_s^4)$ NNLO corrections to the production subprocess are
denoted by $d\tilde{\sigma}^{(1)}_{t\bar{t}}$ and
$d\tilde{\sigma}^{(2)}_{t\bar{t}}$ respectively. They can be obtained
from integrating either the 1PI or PIM hard-scattering kernels. We
point out that the differential cross sections,
$d\sigma^{(0)}_{t\bar{t}}$ and $d\tilde{\sigma}^{(1,2)}_{t\bar{t}}$
receive contributions from $q\bar{q}$ and $gg$ initial states only,
while $d\sigma^{(1)}_{t\bar{t}}$ additionally receives a contribution
from the $qg$ channel.

When including the NLO
corrections to the top or anti-top decays as well as in the
production, the width of the top ought to be treated in a way such that, 
when integrating over the full phase space of the top decay
products, the inclusive cross section for the production of a
$t\bar{t}$-pair multiplied by a branching fraction is recovered. At
NLO and nNLO we adopt the treatment detailed in
\cite{Melnikov:2009dn,Campbell:2012uf} and expand the NLO top width,
keeping only terms in the differential cross section that constitute a
strict NLO correction. This means that in the above equations we
replace $\Gamma^{\text{NLO}}_t \to \Gamma^{\text{LO}}_t$ in the
prefactor and at the same time modify the Born-level differential
cross section by
\begin{align}
d\sigma^{(0)}_{t\bar{t}} \to
d\sigma^{(0)}_{t\bar{t}} \left( 1-2\, 
\frac{\Gamma^{\text{NLO}}_t-\Gamma^{\text{LO}}_t}{\Gamma^{\text{LO}}_t} \right),
\end{align} 
where the factor of two comes from the fact that we have two top
quarks decaying.

\subsection{Checks}

We have performed extensive checks on our results and MC
implementations.  Firstly, our newly computed hard functions for the
production process are consistent with the hard functions computed in
\cite{Ahrens:2010zv}. In addition, by tracing in colour space the
matrix multiplications of the one-loop hard functions with the
corresponding tree-level soft functions we fully reproduce the
numerical results for the one-loop matrix elements, as expected.
Secondly, regarding our implementation of the approximate (N)NLO
kernels into a MC and the associated phase-space integration, when we
are inclusive over the top decay products (i.e. no analysis cuts
applied), we find complete agreement between the numbers produced by
our code and n(N)LO inclusive cross sections results presented in
\cite{Ahrens:2010zv, Ahrens:2011mw, Ahrens:2011px}.  Furthermore,
again when we integrate over the top decay products, we fully
reproduce the differential distributions presented in
\cite{Ahrens:2010zv, Ahrens:2011mw}.  This is a strong check
indicating that our MC implementation of the n(N)LO contributions is
correct, both for PIM and 1PI kinematics.  Finally, regarding the
treatment of the top width, we have compared several observables for
the exclusive setup of \Eqn{eq:cut-def} with the publicly available
code MCFM \cite{Campbell:2012uf}, with which we find agreement within
MC errors (better than 1\%) at NLO, both when NLO corrections are or
are not included in the decay. This provides an excellent check of our
implementation of the NLO corrections, both in production and decay as
well as our treatment of the top width.

\section{Results}
\label{sec:res}

\subsection{Total cross sections}
\label{sec:resTotXS}

We first assess the quality of the approximation by analyzing the
$t\bar{t}$ total cross sections. We compare the n(N)LO vs the (N)NLO
results taking into account the scale uncertainties. The NNLO cross
sections are obtained using {\tt Top++} \cite{Czakon:2013goa}.  This
analysis is not meant to be a systematic study of the $t\bar{t}$ total
cross sections (for detailed studies we refer to
\cite{Czakon:2013goa,Ahrens:2011px}) but a validation of the
method. Throughout this subsection we use MSTW08NLO PDFs
\cite{Martin:2009iq} and a top mass value of $m_t = 173.3$ GeV. The
scale variation of the cross sections is obtained by setting the
factorization and the renormalization scales equal to each other,
$\mu_f = \mu_R = \mu$, and by looking at the predictions for different
values of this single scale, in particular we choose $\mu =
\{m_t/2,m_t, 2 m_t\}$.  The total cross sections are obtained by
integrating the hard-scattering kernels over the complete (NLO, PIM or
1PI) phase space (no cuts on final state). The cross sections obtained
with our code are reported in Table~\ref{tab:totalcs}.

\renewcommand{\arraystretch}{1.15}
\begin{table}[th!]
\centering
\begin{tabular}{|c||c|c|c|}
\hline
 & LHC 8 TeV & LHC 14 TeV & Tevatron \\
\hline
\hline 
$(\sigma\pm \Delta\sigma_\mu)_{\rm NLO}$ [pb] & ${226}^{+28}_{-29}$ &
${884}^{+107}_{-106}$ &${6.68}^{+0.36}_{-0.75}$ \\
$(\sigma\pm \Delta\sigma_\mu)_{{\rm PIM}_{\rm nLO}}$ [pb] &${216}^{+8}_{-20}$ &
${831}^{+18}_{-60}$ & ${6.38}^{+0.42}_{-0.76}$\\
$(\sigma\pm \Delta\sigma_\mu)_{{\rm 1PI}_{\rm nLO}}$ [pb] &${231}^{+0}_{-14}$ &
${883}^{+0}_{-65}$ & ${6.75}^{+0.19}_{-0.70}$ \\
\hline
$(\sigma\pm \Delta\sigma_\mu)_{\rm NNLO}$ [{\tt Top++}] [pb]& 
${254}^{+10}_{-17}$ & ${989}^{+36}_{-57}$& ${7.29}^{+0.29}_{-0.41}$ \\
$(\sigma\pm \Delta\sigma_\mu)_{{\rm PIM}_{\rm nNLO}}$ [pb] & 
${233}^{+13}_{-12}$ &${905}^{+51}_{-47}$ & ${6.82}^{+0.05}_{-0.43}$ \\
$(\sigma\pm \Delta\sigma_\mu)_{{\rm 1PI}_{\rm nNLO}}$ [pb] & 
${234}^{+6}_{-3}$ &${894}^{+27}_{-7}$ & ${6.85}^{+0}_{-0.30}$ \\
\hline
\end{tabular}
\caption{Total cross sections at the LHC 8 TeV, LHC 14 TeV and the
  Tevatron for a top mass $m_t = 173.3$ GeV. The theoretical
  uncertainty is obtained by varying the scale in the range $\mu =
  \{m_t/2,m_t, 2 m_t\}$. The scale choice $\mu = m_t$ is taken as the
  central value prediction. The NNLO cross sections are produced with
  the program {\tt Top++} \cite{Czakon:2013goa}.  All the cross
  sections in the table are produced with MSTW08NLO PDFs.}
\label{tab:totalcs}
\end{table}

\renewcommand{\arraystretch}{1.0}
\begin{figure}[t!]
\centering
\begin{center}
\begin{tabular}{ccc}
  \includegraphics[width=0.31\textwidth]{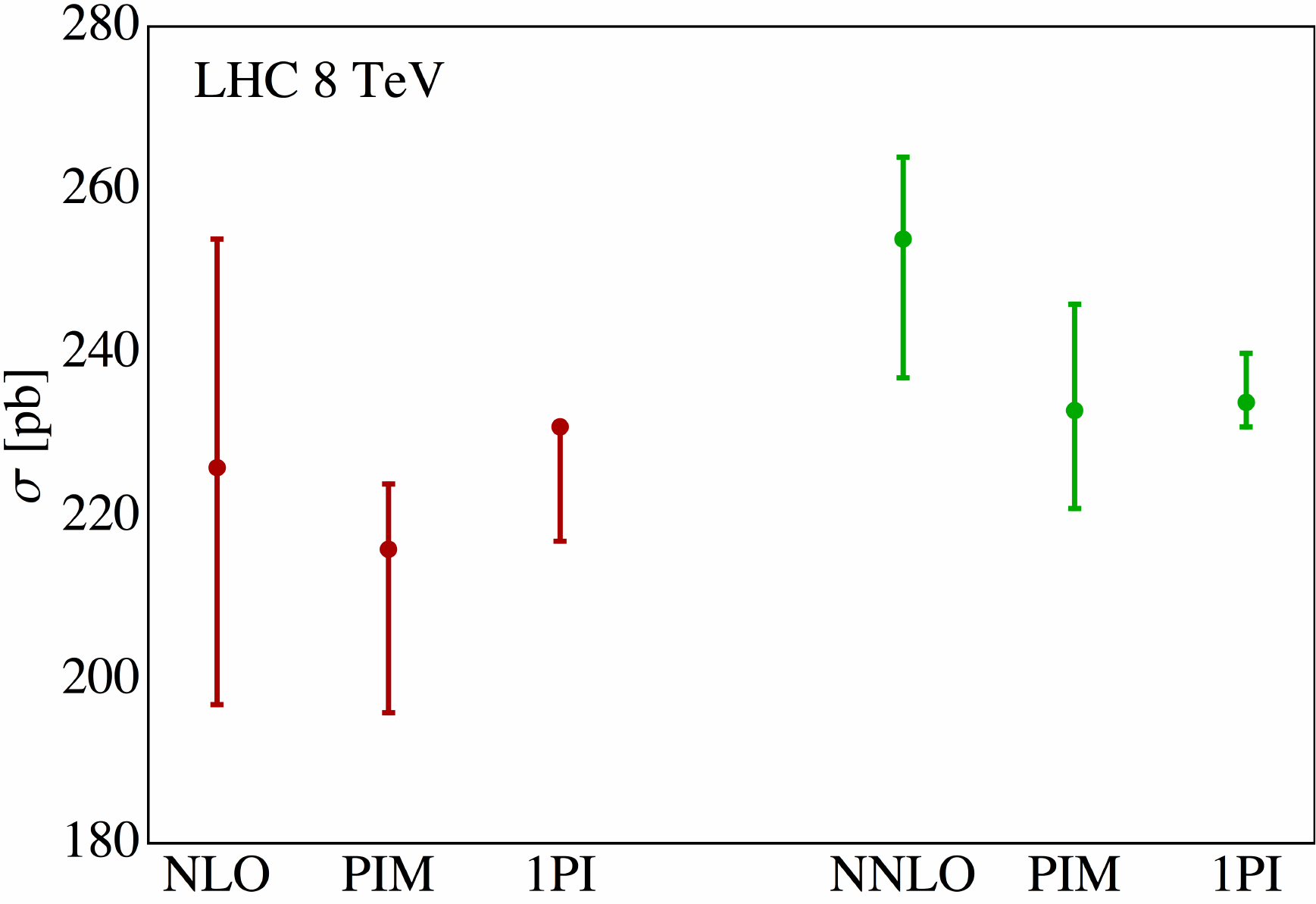} & 
  \includegraphics[width=0.318\textwidth]{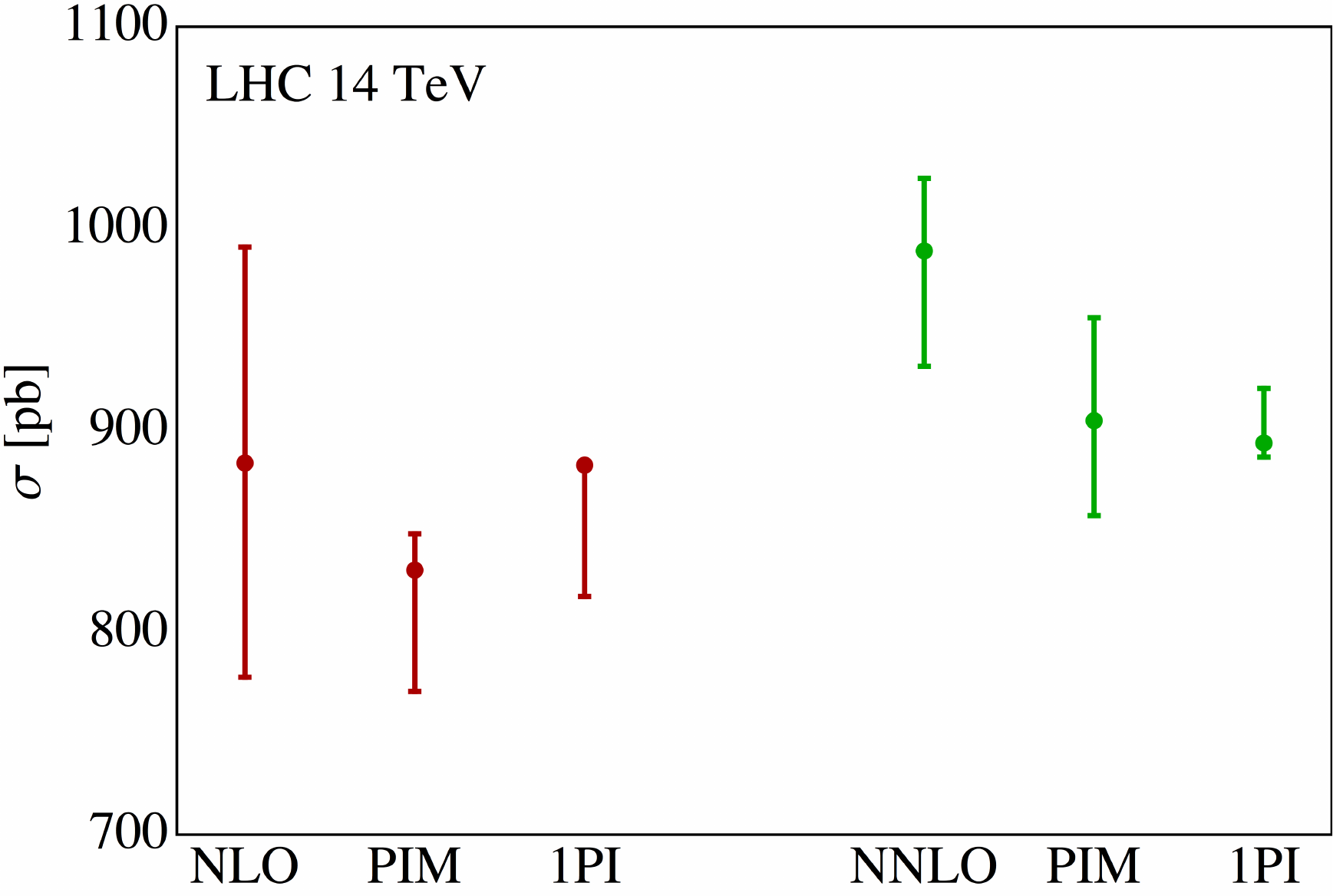} &
  \includegraphics[width=0.31\textwidth]{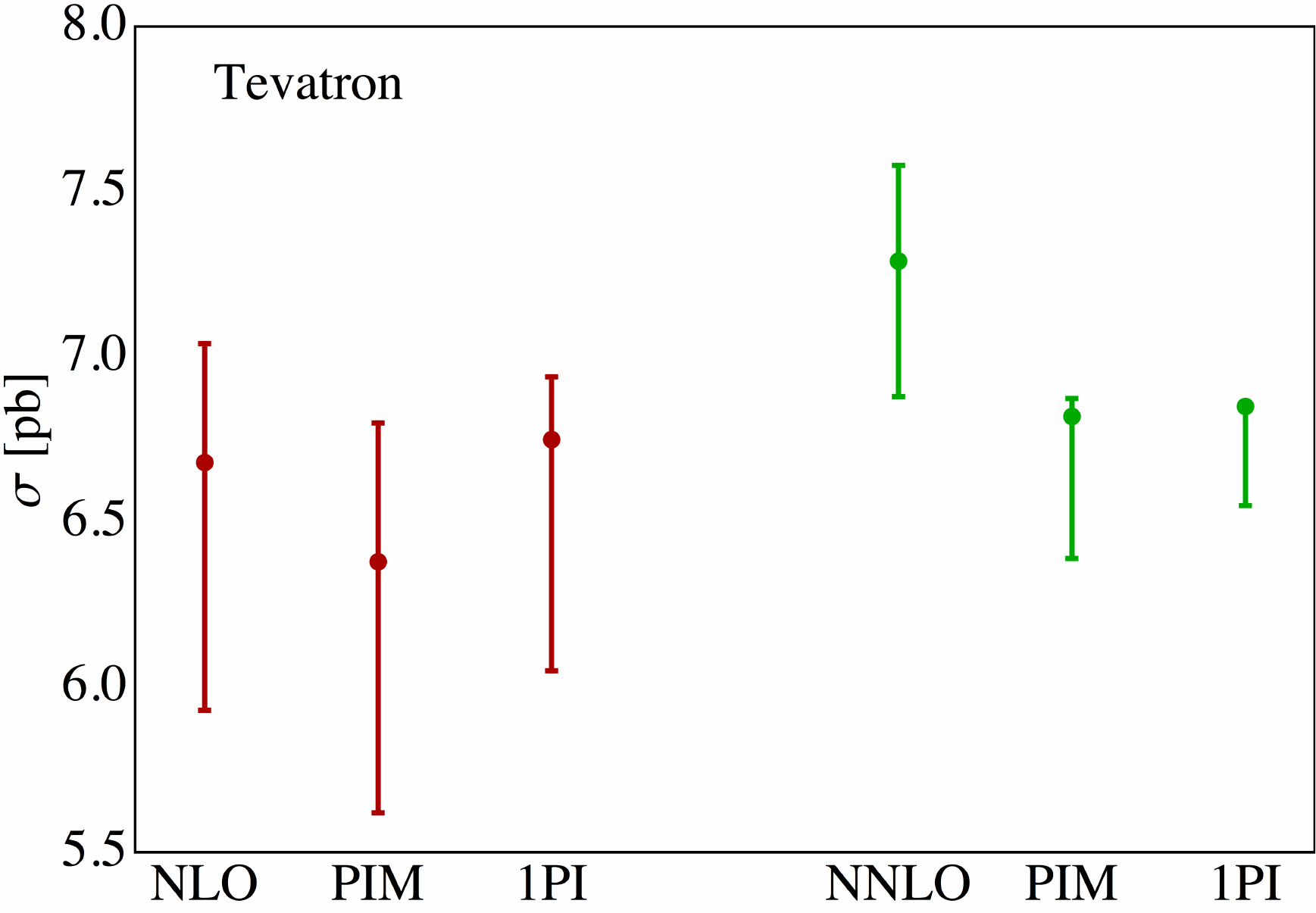} \\
\end{tabular} 
\end{center}
\label{fig:totcs} 
\caption{Graphical representation of the cross sections reported in
  Table~\ref{tab:totalcs} including the error bands for the
  theoretical scale uncertainty. We show the results for the LHC 8
  TeV, LHC 14 TeV and the Tevatron. The red bands refer to the
  NLO and nLO results while the green bands refer to the NNLO and
  nNLO results. The nNLO cross sections are matched with fixed-order
  NLO calculations.}
\end{figure}

In Figure~\ref{fig:totcs} we show a graphical representation of the
cross sections reported in Table~\ref{tab:totalcs} including the error
bands for the scale uncertainty.  The red bands refer to the NLO
results and the nLO results in PIM and 1PI kinematics.  At nLO, both
for PIM and 1PI, the central values are very close to the NLO results
and are consistent with each other. The individual PIM and 1PI nLO
error bands underestimate the theoretical uncertainty of the NLO
results, in particular for the LHC. For this reason we choose to
provide a more realistic estimate of the theoretical uncertainty by
taking the envelope of the PIM and 1PI bands. We should also mention
that the nLO results do not include the $qg$ channel. Therefore a more
direct comparison between nLO and NLO could be made individually for
the $q\bar{q}$ and $gg$ channels.  We refer to the next subsections
where this comparison is made at the level of the distributions.

The green bands refer to the NNLO results and to the nNLO results in
PIM and 1PI kinematics.  The PIM and 1PI nNLO central values are very
close and consistent with each other independently of the
collider. Unfortunately, the central values of both PIM and 1PI 
fall below the NNLO central cross sections. Nevertheless, at the LHC
there is quite a large overlap between the NNLO lower uncertainty
bands and the nNLO upper bands. At the Tevatron however the overlap is
only marginal.  The fact that the nNLO predictions are consistently
lower than the NNLO ones is likely due to the fact that a large part
of the corrections that contribute to the delta terms in
\Eqns{eq:pimkernels}{eq:1pikernels} is missing. In particular, as
mentioned in Section~\ref{subsec:structurehsk}, the finite
contributions of the two-loop hard and soft functions are not included
in our nNLO result. Comparing to the nLO predictions, where the
delta terms are fully known, we observe a better agreement between nLO
and NLO. When the delta contributions become available we expect that
the comparison of nNLO vs NNLO will be as good as the nLO vs NLO
case. The discrepancy between the approximate and full results may
also partly originate from the phase space. Integrating the PIM and
1PI kernels over a full phase space instead of an approximate one
could lead to a positive contribution to the cross sections. We do not
investigate this second possibility here, but postpone this study to
future work.

We deliberately refrain from adapting the treatment of subleading
terms to optimize the agreement between the nNLO and NNLO results for
the total cross section, because such an a posteriori justification
for a particular treatment of subleading terms cannot necessarily be
generalized to arbitrary distributions.  From the comparison between
the nNLO and NNLO results for the total cross section we conclude that
our differential distributions are likely to be at the lower end of
the NNLO predictions for the same observables. However, when the fully
differential NNLO results will be available, we do expect to find a
substantial overlap between the uncertainty bands of the nNLO and NNLO
distributions.

\subsection{Universality and uncertainties of approximate results}
\label{sec:consistency}

In this subsection we study the reliability of our method for
distributions. To do so, we compare our (approximate) nLO results to
full NLO results with the aim of obtaining a procedure that gives a
reliable error estimate for our approximate results.

We will use the PIM and/or the 1PI kinematics to obtain approximate
expressions for arbitrary observables. Thus, we have to investigate
the universality of the approximate terms obtained using the two
different kinematics. Furthermore, the impact of cuts that might be
applied in a realistic analysis has to be considered as well.

To get an idea of the latter, we will apply a jet algorithm and
require that the event has a $b$-jet, $J_b$ and a $\bar{b}$-jet,
$J_{\bar{b}}$.  As an example we have used the $k_T$ clustering algorithm
with the resolution parameter set to $R=0.7$, but obviously any other
jet definition would be possible.  All observables we study will be
constructed from the momenta of the final-state objects $J_b$,
$J_{\bar{b}}$ and the decay products of the $W$ bosons. We will only
consider the decay $W^+ \to e^+\, \nu$ and $W^- \to e^-\, \bar{\nu}$
and sometimes assume the $W$ bosons can be reconstructed fully from
their decay products. In particular, the momentum of the reconstucted
top is defined as $p(t) \equiv p(W^+)+p(J_b) \neq p_t$, with an
analogous expression for the reconstructed anti-top. We will also
study the impact of applying some additional standard cuts on the
transverse momenta $p_T$, and transverse (missing) energies $E_T$. The
precise definition of these cuts is as follows
\begin{align}
p_T(J_b) > 15~{\rm GeV} \qquad &
p_T(J_{\bar{b}}) > 15~{\rm GeV}  \qquad 
M(W^+,W^-,J_b,J_{\bar{b}}) > 350~{\rm GeV} \label{eq:cut-def}\\
E_T(e^+) > 15~{\rm GeV} \qquad &
E_T(e^-) > 15~{\rm GeV} \qquad
\slash{E}_T > 20~{\rm GeV} \nonumber
\end{align}
These cuts are meant as an illustration only. They can easily be
changed and, in particular, additional cuts on the rapidity can be
applied.

\begin{figure}[t!]
\centering
\begin{tabular}{c c}
\includegraphics[trim=0.5cm 0.6cm 0.7cm
  0.8cm,clip,width=0.48\textwidth]{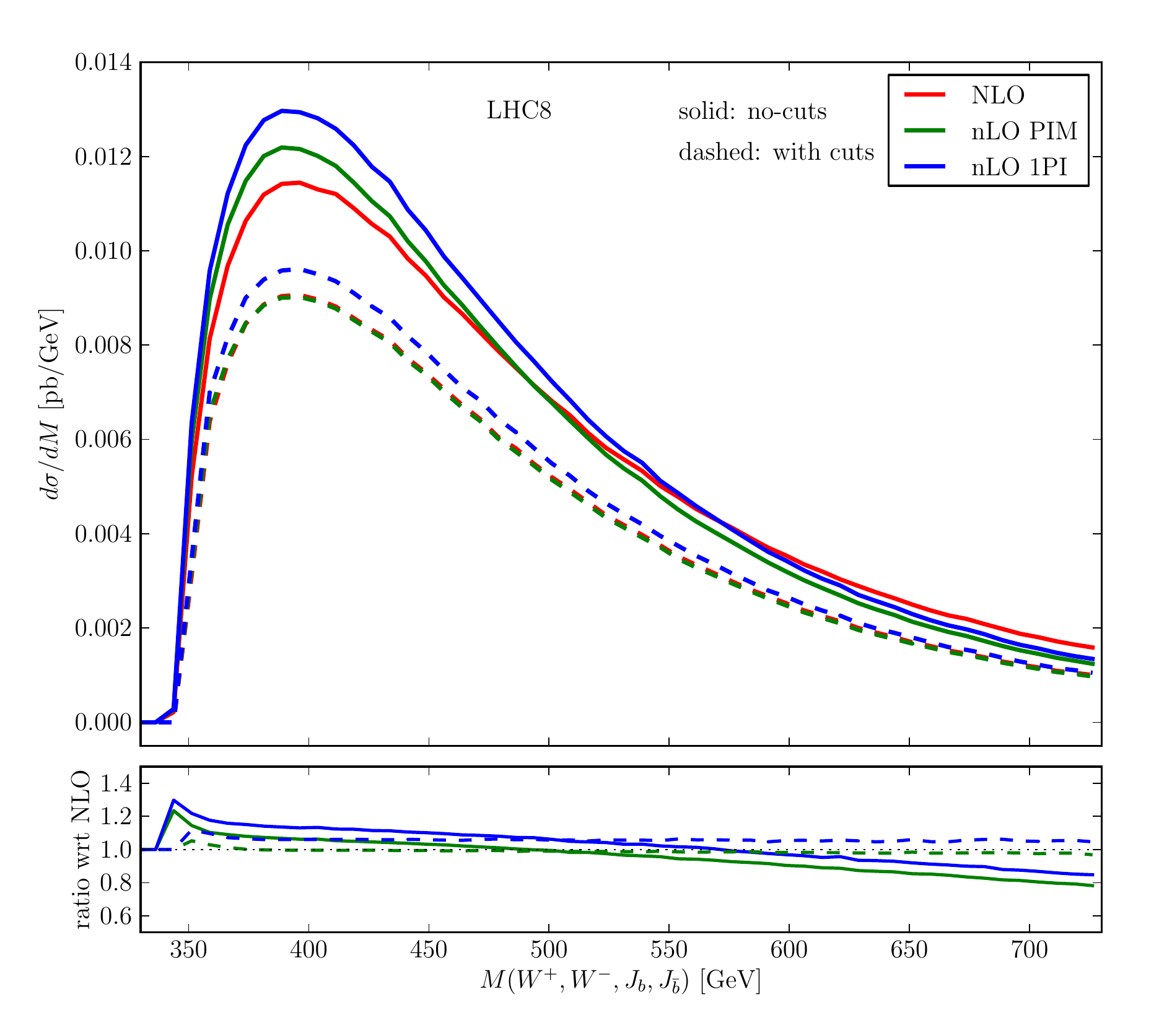} &
\includegraphics[trim=0.5cm 0.6cm 0.7cm
  0.8cm,clip,width=0.48\textwidth]{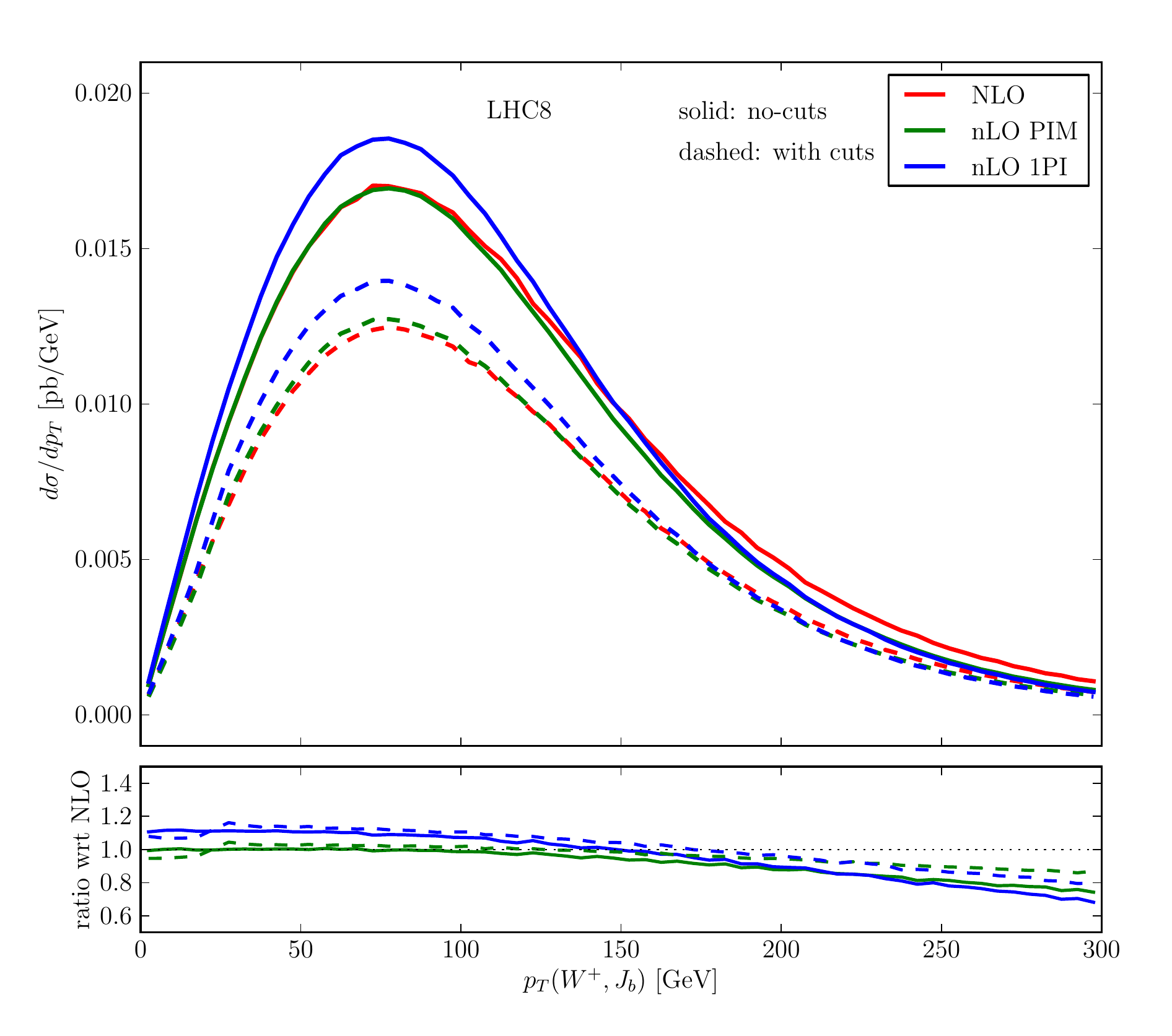} 
\end{tabular}
\caption{Invariant mass of reconstructed top-pair system (left) and
  transverse momentum of reconstructed top quark (right) for
  LHC8. Results without (with) analysis cuts shown as solid (dashed)
  lines. Production corrections only. }
\label{fig:log-checks}
\end{figure}

To study the dependence of the predictions on the kinematics applied,
we start by considering in the left panel of
Figure~\ref{fig:log-checks} the distribution of the invariant mass of
the reconstructed top pair $M(t,\bar{t}) \equiv M(W^+, W^-, J_b,
J_{\bar{b}})$. The appropriate kinematics for this observable is PIM,
but we will also use the `wrong' kinematics 1PI. We want to
investigate how well the approximate nLO result reproduces the full
NLO corrections to the production of the top pair. Note that the NLO
corrections to the decay are not affected and, hence, will be left out
in this comparison. Thus, in Figure~\ref{fig:log-checks} we show the
comparison of the nLO results $d\sigma^{\text{nLO}}_{\text{prod}}$
obtained with PIM (green) and 1PI (blue) to the full NLO result
$d\sigma^{\text{NLO}}_{\text{prod}}$ (red). In both cases the decay
of the (anti-)top is included at LO only and we are using
MSTW08NLO PDFs~\cite{Martin:2009iq} and set
$\mu_f = \mu_R = m_t$.

The main message is that the approximate results for both kinematics
are in reasonably good agreement with the exact results and the
application of cuts does not have a negative effect on this
agreement. Actually in this case it turns out that the agreement is
even better if cuts are applied (dashed curves).

As expected, for the invariant mass of the top pair the PIM results
are slightly better than the 1PI results. We now repeat this
comparison for the transverse momentum distribution of the
reconstructed top quark $p_T(t) \equiv p_T(W^+, J_b)$, where 1PI is
the appropriate kinematics. In the right panel of
Figure~\ref{fig:log-checks} we show again the comparison of
$d\sigma^{\text{nLO}}_{\text{prod}}$ obtained with 1PI and PIM to
$d\sigma^{\text{NLO}}_{\text{prod}}$. Once more, the approximations
using both kinematics work well, with and without applying cuts. What
is somewhat surprising is that the `wrong' kinematics, PIM, gives
better agreement with the full NLO results than the `right'
kinematics, 1PI. We have also investigated the rapidity of the top in
an analogous manner. Again, both PIM and 1PI give very good
approximations. If no cuts are applied, the 1PI result gives better
agreement as one would expect. However PIM does slightly better than
1PI if cuts are applied.

To summarize these considerations we can state that the bulk of the
NLO corrections to the production obtained using approximate results
appears to be almost independent of the kinematics applied. The
difference between PIM and 1PI results are of course due to the fact
that different approximations have been made in the two kinematics. As
for the total cross section, this difference has to be taken into
account in addition to the usual scale dependence when giving a
theoretical error for the approximate result for a generic observable.

\begin{figure}[t!]
\centering
\begin{tabular}{c c}
\includegraphics[trim=0.3cm 0.6cm 0.7cm
  0.4cm,clip,width=0.48\textwidth]{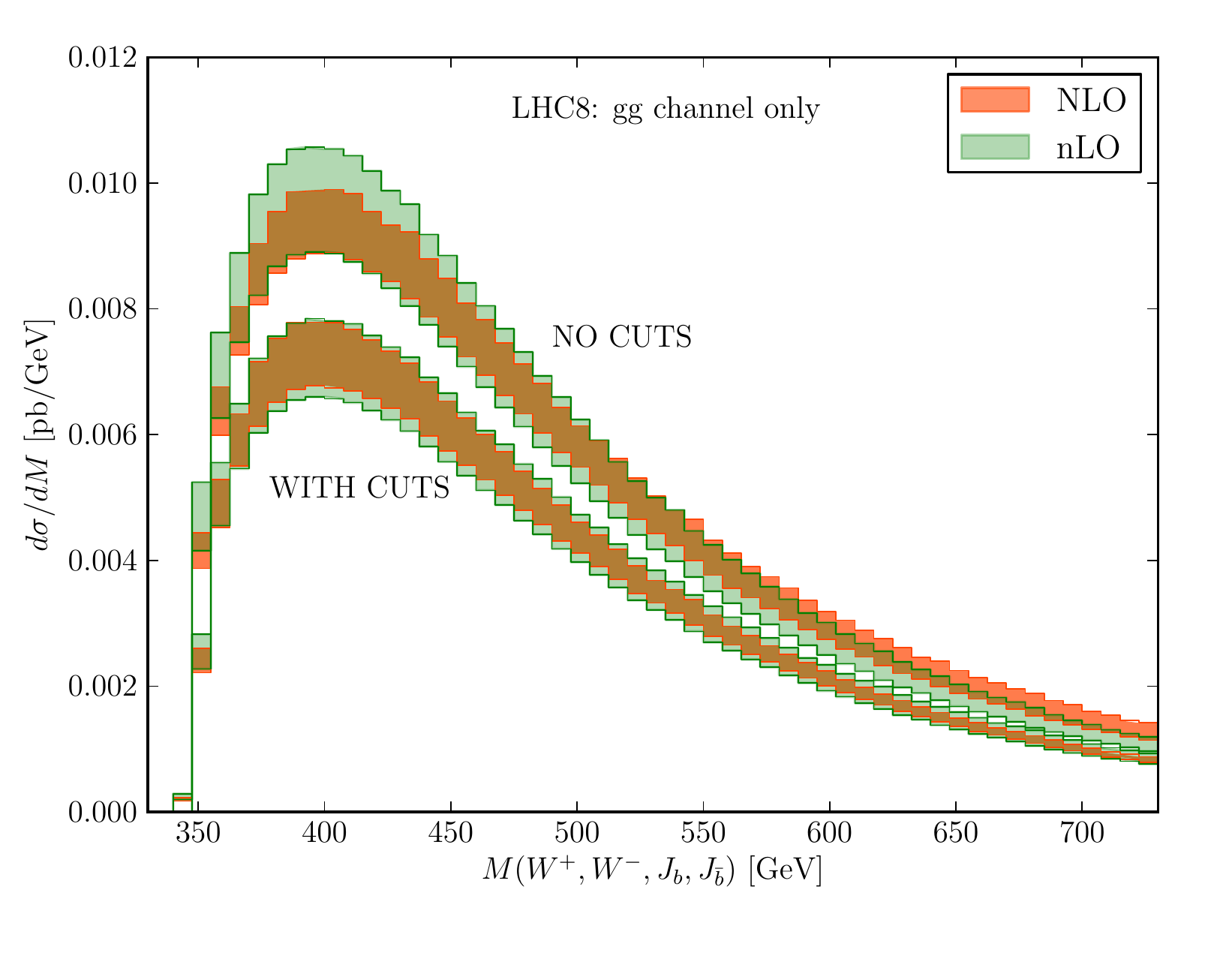} &
\includegraphics[trim=0.3cm 0.6cm 0.7cm
  0.4cm,clip,width=0.48\textwidth]{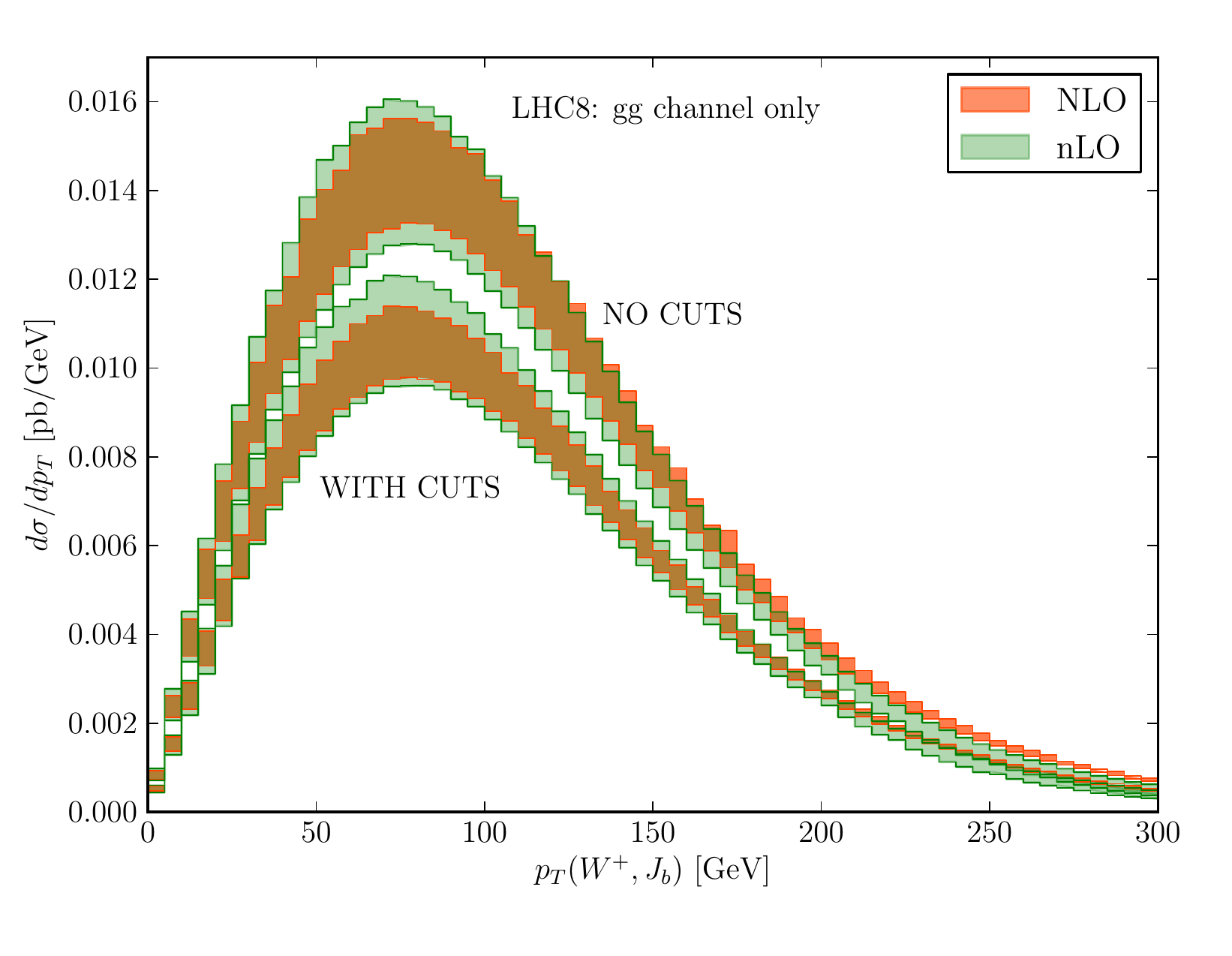} 
\end{tabular}
\caption{Invariant mass of reconstructed top-pair system (left) and
  transverse momentum of reconstructed top quark (right) for $gg$
  partonic channel at LHC8, without cuts (upper bands) and with
  analysis cuts (lower bands). Production corrections only. }
\label{fig:uncert-bands}
\end{figure}

In order to obtain a procedure for assigning a realistic error estimate
of our approximate results, we start by considering a single partonic
channel, the dominant $g g$ initial state for the LHC at 8~TeV. Again
we consider corrections to the production only and will take
$M(t,\bar{t})$ (left panel of Figure~\ref{fig:uncert-bands}) and
$p_T(t)$ (right panel of Figure~\ref{fig:uncert-bands}) as our test
observables.  For the complete NLO result we use the conventional
scale variation $m_t/2 \le \mu \le 2 m_t$ and show the uncertainty
as a red band in Figure~\ref{fig:uncert-bands}. The upper (lower)
bands are without (with) our standard cuts. The uncertainty band for
our approximate results are obtained by taking the envelope of the
scale variation and the variation over the implementations.  These
bands are shown as green bands in Figure~\ref{fig:uncert-bands}. 

Looking at the results shown in Figure~\ref{fig:uncert-bands} we see
that for the bulk of the cross section, the bands obtained with our
error estimate for the nLO results contains the error band of the full
NLO result. This is the case whether or not we apply cuts. At the
high-end tail of the $p_T$ distribution, the agreement between the
full and approximate results gets somewhat worse. This is not
surprising, as this region is outside the range of applicability of
the approximations we make. However, this region will not have a large
effect on generic distributions. We also note that it seems to be a
common feature that the application of cuts improves the quality of
the agreement between the full and approximate results.

Generally speaking, our procedure overestimates the error of the
approximate results for a particular partonic channel. On the other
hand, the approximate nLO results for an actual observable do not
include all partonic channels. In fact, as mentioned previously,
the nLO results do not include
any contributions at all from the $gq$ channels. Thus the overestimate
of the uncertainty in the partonic channel $gg$ (and $q\bar{q}$) can
be seen as a compensation for the complete omission of partonic
channels that are not included in the approximate results.

\begin{figure}[t!]
\centering
\begin{tabular}{c c}
\hspace{-0.5cm}
\includegraphics[trim=0.3cm 0.6cm 0.7cm 0.4cm,clip,width=0.48\textwidth]{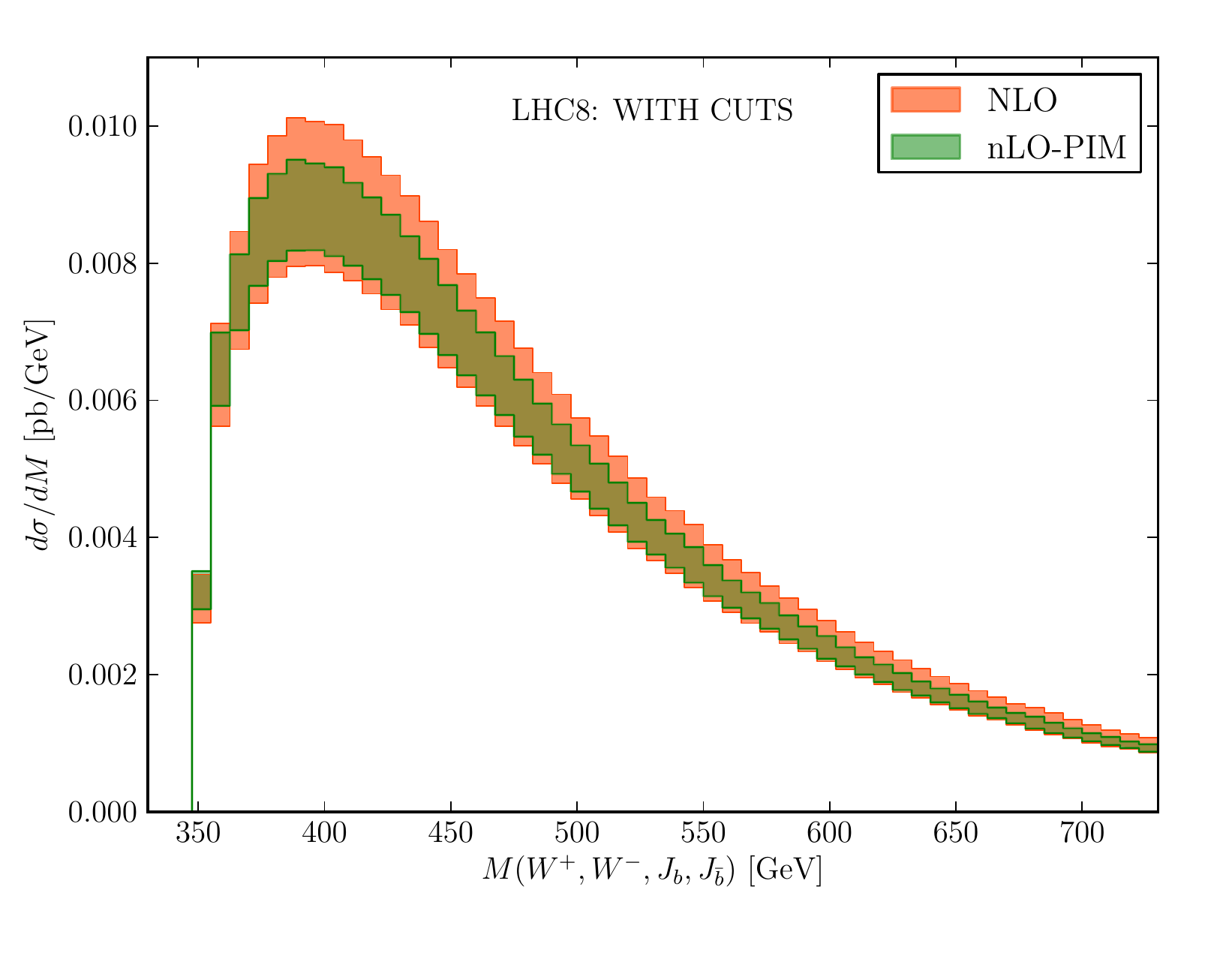} &
\hspace{-0.6cm}
\includegraphics[trim=0.3cm 0.6cm 0.7cm 0.4cm,clip,width=0.48\textwidth]{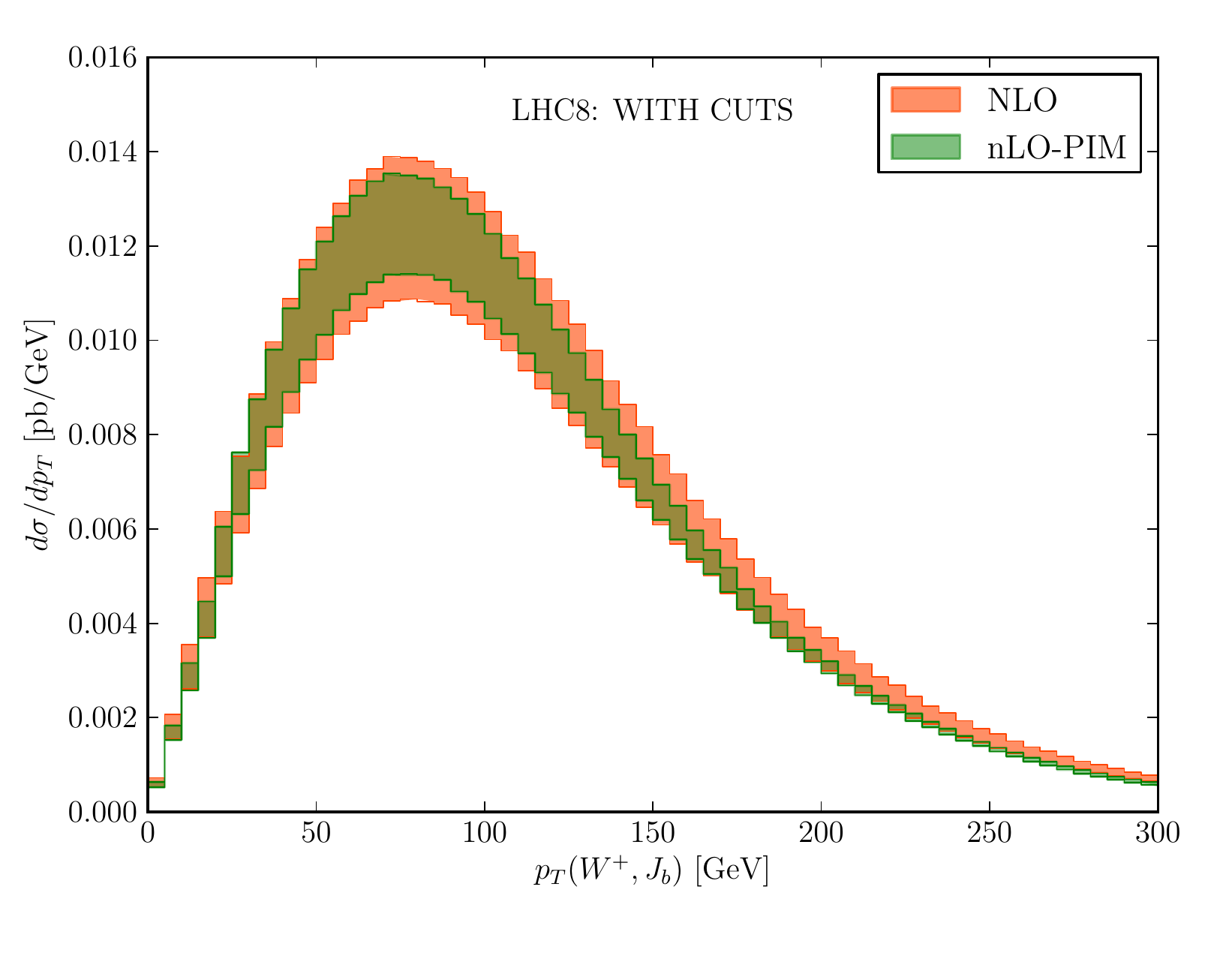} \\ 
\hspace{-0.5cm}
\includegraphics[trim=0.3cm 0.6cm 0.7cm 0.4cm,clip,width=0.48\textwidth]{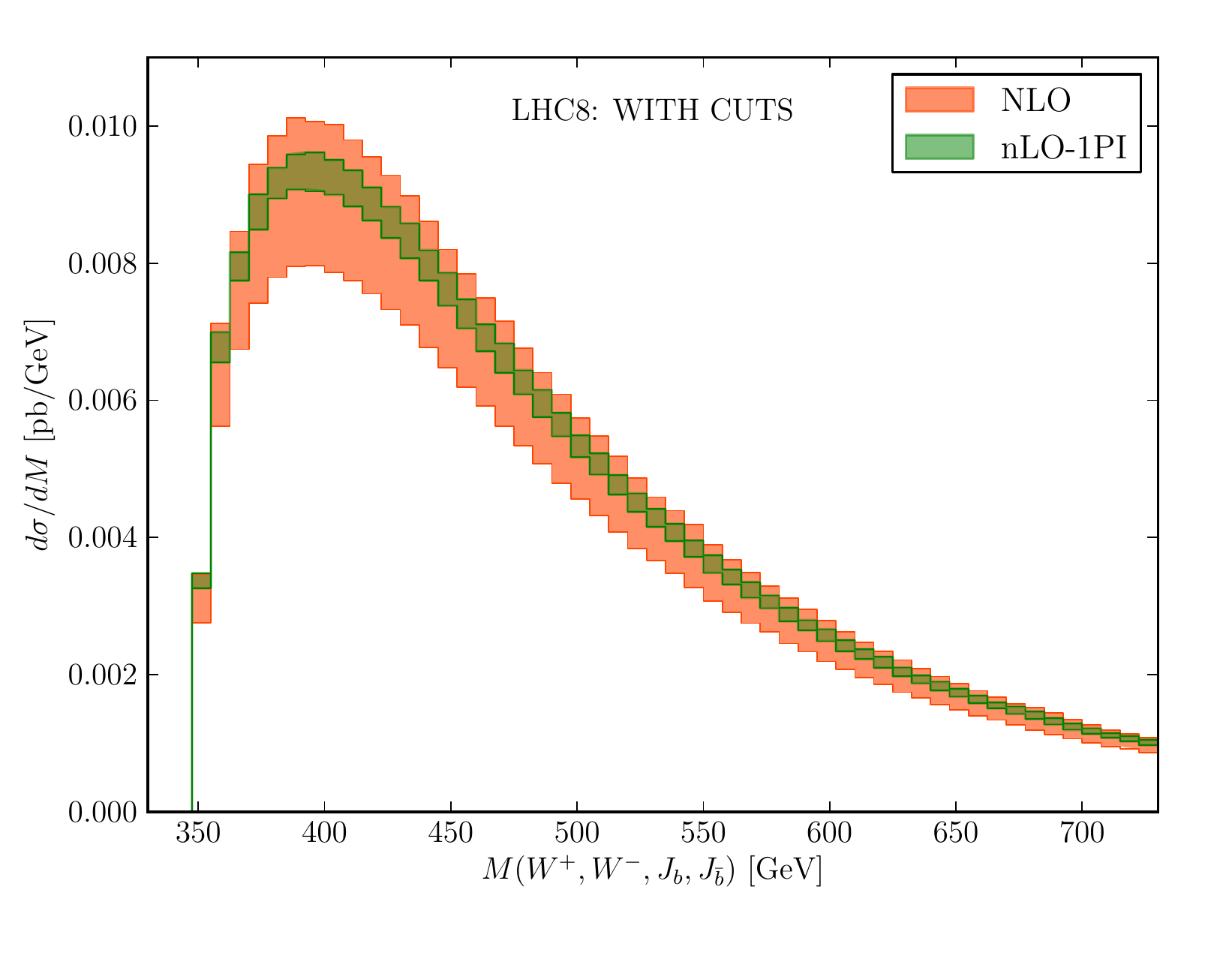} &
\hspace{-0.5cm}
\includegraphics[trim=0.3cm 0.6cm 0.7cm 0.4cm,clip,width=0.48\textwidth]{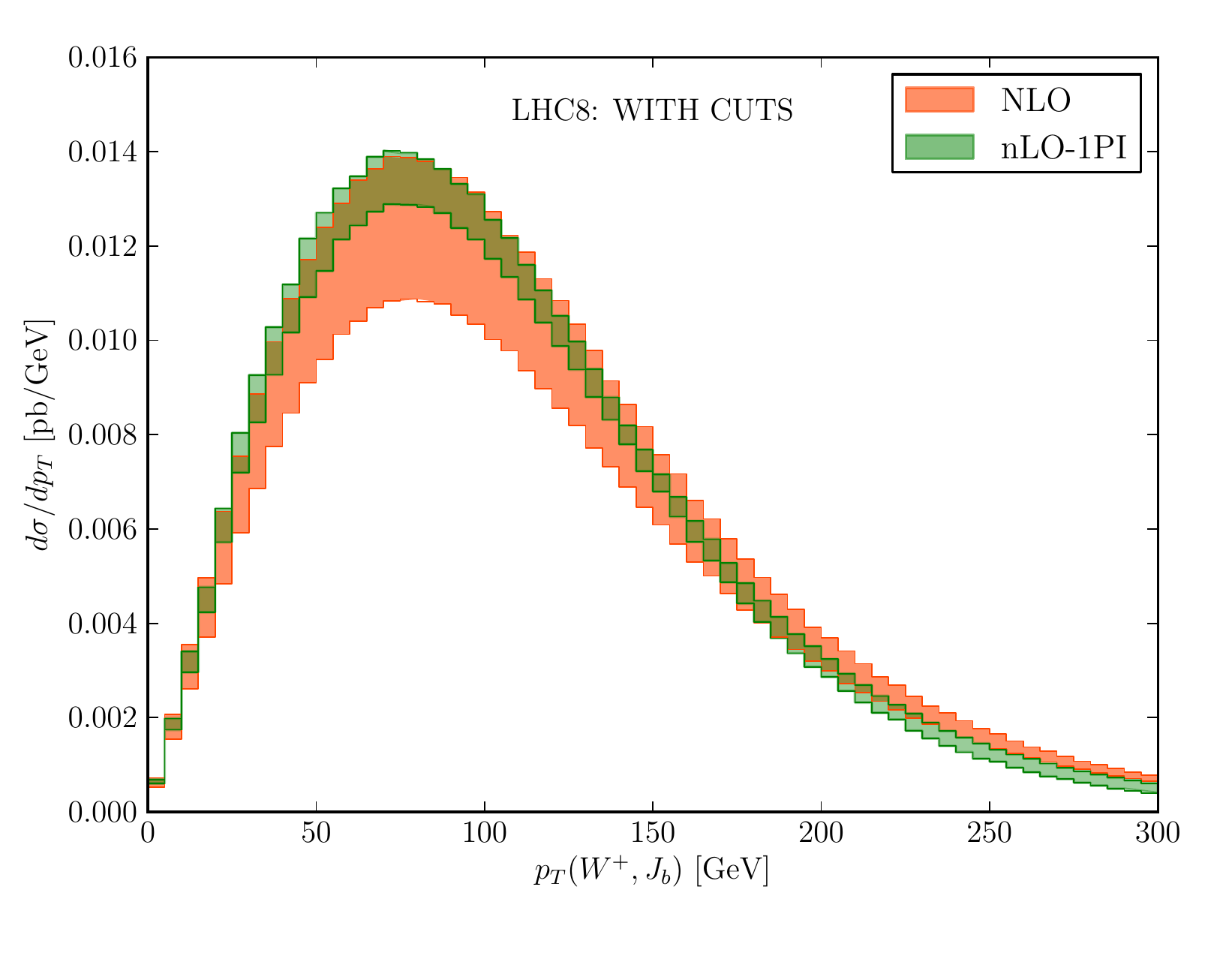} \\ 
\end{tabular}
\caption{$M(W^+,W^-,J_b,J_{\bar{b}})$ (left) and $p_T(W^+,J_b)$
  (right) distributions for LHC, with analysis cuts. The NLO uncertainty
  band is shown in all plots in red. The green bands are the nLO
  predictions, with PIM predictions shown in the top two plots whilst
  those of 1PI shown in the bottom two. }
\label{fig:pT-Mtt-bands-quality}
\end{figure}

We would like to stress that a reliable error estimate requires taking
the envelope over the different kinematics. Simply using either PIM or
1PI, and considering only the scale variation usually
underestimates the error. This is illustrated in
Figure~\ref{fig:pT-Mtt-bands-quality}, where we show a comparison of
the standard NLO bands with nLO bands obtained for a fixed
kinematics. Contrary to Figure~\ref{fig:uncert-bands}, in
Figure~\ref{fig:pT-Mtt-bands-quality} we include all partonic
channels. The top panel shows the error obtained from using PIM and
taking the envelope from scale variation only, whereas in the bottom
panel 1PI only is used. It is clear that in particular the 1PI bands
obtained solely from scale variation considerably underestimate the
theoretical error. However, if we take the envelope over both
kinematic implementations as well as the scale dependence, we get a
more realistic indication of the uncertainty.

We have verified that this procedure works for generic observables in
that the error band obtained this way has a large overlap with
the standard error band from scale variation of the full NLO
result. This is consistent with what we have found in
Section~\ref{sec:resTotXS} for the total cross section.

\subsection{Differential distributions for LHC8, LHC14 and Tevatron}
\label{subsec:collider-diff-results}

We will now consider generic distributions at different colliders
where the decay of the tops is included at NLO and where we always
apply our standard cuts \Eqn{eq:cut-def}.  Our best predictions will
be $d\sigma^{\text{nNLO}}_{\text{full}}$ as defined
in~\Eqn{eq:nNLO-full}.  This result contains the exact NLO corrections
for the production and the decay and approximate NNLO corrections to
the production. For LO, nLO and NLO results MSTW08NLO PDFs are used
while MSTW08NNLO PDFs are used for the nNLO results.  The band of
these results (shown in green in the figures of this section) is
obtained by taking the envelope of the standard scale dependence
$m_t/2 \le \mu \le 2 m_t$ \emph{and} the variation over PIM and 1PI
kinematics. We have seen in Section~\ref{sec:resTotXS} that when using
this procedure for the LHC, the total cross section gives bands that
have an overlap with the full NNLO bands (obtained from standard scale
variation) but are generally somewhat  lower. For the Tevatron, the
overlap is only marginal. Thus a certain care has to be taken when
interpreting these bands as theoretical error.

\begin{figure}[t!]
\centering
\begin{tabular}{c c}
\includegraphics[trim=0.3cm 0.6cm 0.7cm
  0.4cm,clip,width=0.48\textwidth]{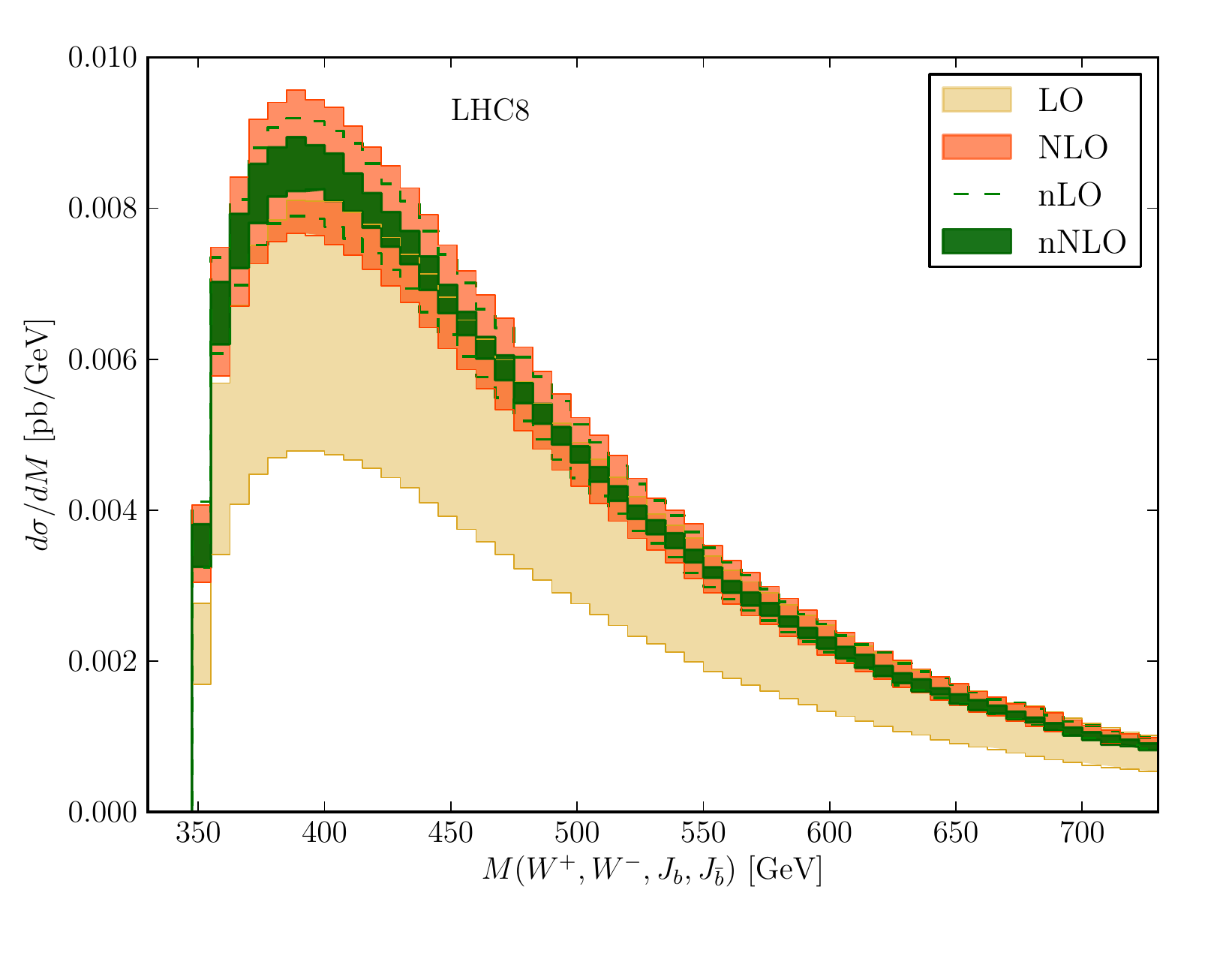} &
\includegraphics[trim=0.3cm 0.6cm 0.7cm
  0.4cm,clip,width=0.48\textwidth]{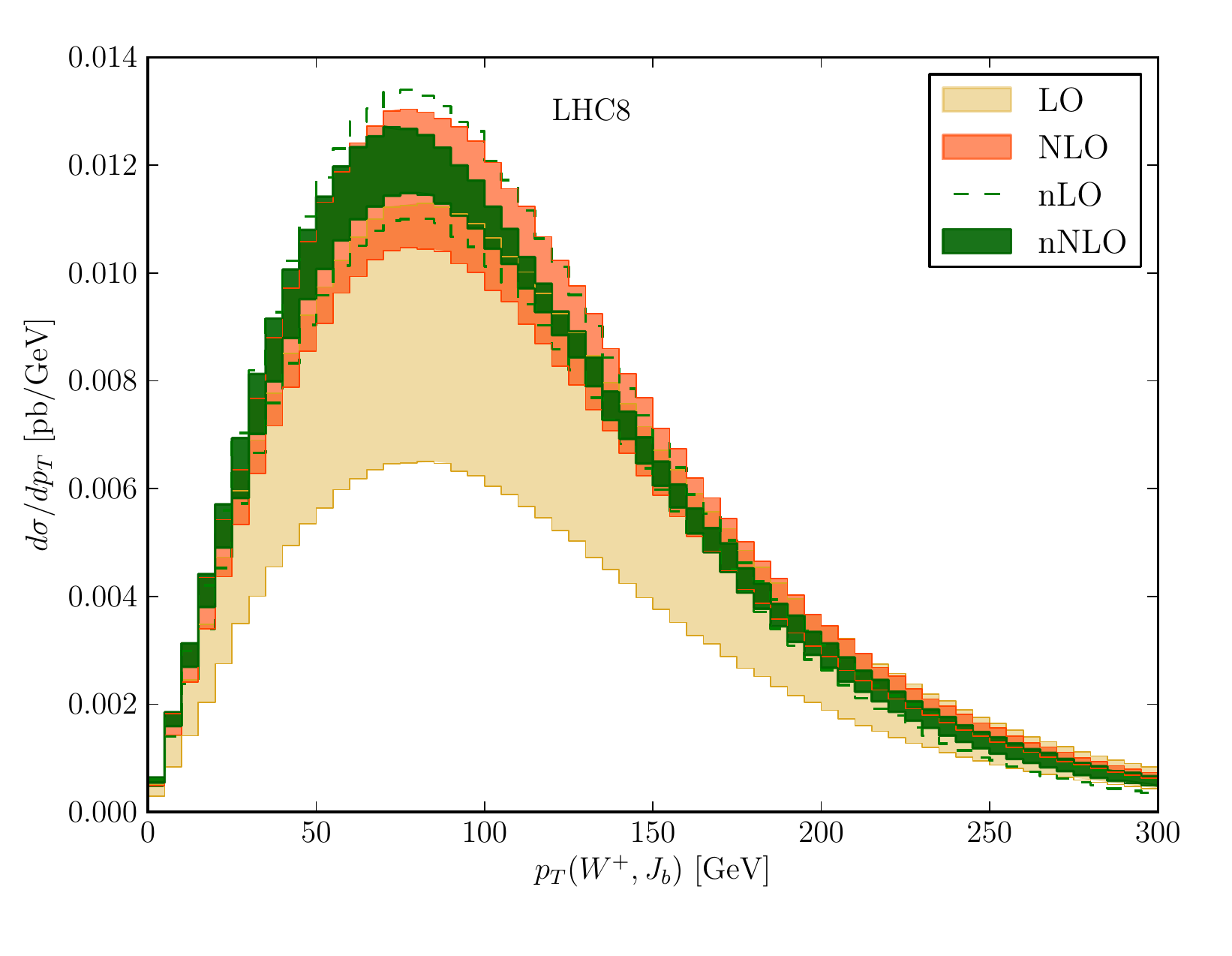} 
\end{tabular}
\caption{Invariant mass of reconstructed top-pair system (left) and
  transverse momentum of reconstructed top quark (right) LHC8 with
  analysis cuts. }
\label{fig:lhc8-mtt-ptt}
\end{figure}

To give a complete picture, we will also show the LO
and NLO results $d\sigma^{\text{NLO}}_{\text{full}}$ as
light brown and red bands respectively, These bands have been obtained
by varying $m_t/2 \le \mu \le 2 m_t$. Finally, in these plots we give
an indication of the quality of the $d\sigma^{\text{nLO}}_{\text{full}}$
approximation. We show the corresponding bands obtained in the same
way as for $d\sigma^{\text{nNLO}}_{\text{full}}$ as green dashed lines. These
bands are to be compared to the red bands to assess the quality of the
approximation at NLO. The quality of the nLO approximation gives some
indirect evidence on the quality of the $d\sigma^{\text{nNLO}}_{\text{full}}$
approximation.

In Figure~\ref{fig:lhc8-mtt-ptt} we present our final results for our
two standard variables $M(W^+,W^-,J_b,J_{\bar{b}})$ and
$p_T(W^+,J_b)$.  As expected the scale dependence is reduced going
from LO to NLO. There is also a very large overlap between the red
$d\sigma^{\text{NLO}}_{\text{full}}$ and green dashed
$d\sigma^{\text{nLO}}_{\text{full}}$ bands. Only for the large $p_T$
tail, where the approximation we make is not justified, the nLO
results start to differ from the NLO results. Finally, the bands of
the nNLO results, $d\sigma^{\text{nNLO}}_{\text{full}}$, are
considerably smaller than and mostly within the NLO bands. This
suggests that the perturbative expansion is under control and there
are no unexpected large corrections in going from NLO to NNLO.

\begin{figure}[t!]
\centering
\begin{tabular}{c c}
\hspace{-0.5cm}
\includegraphics[trim=0.3cm 0.6cm 0.7cm 0.4cm,clip,width=0.48\textwidth]{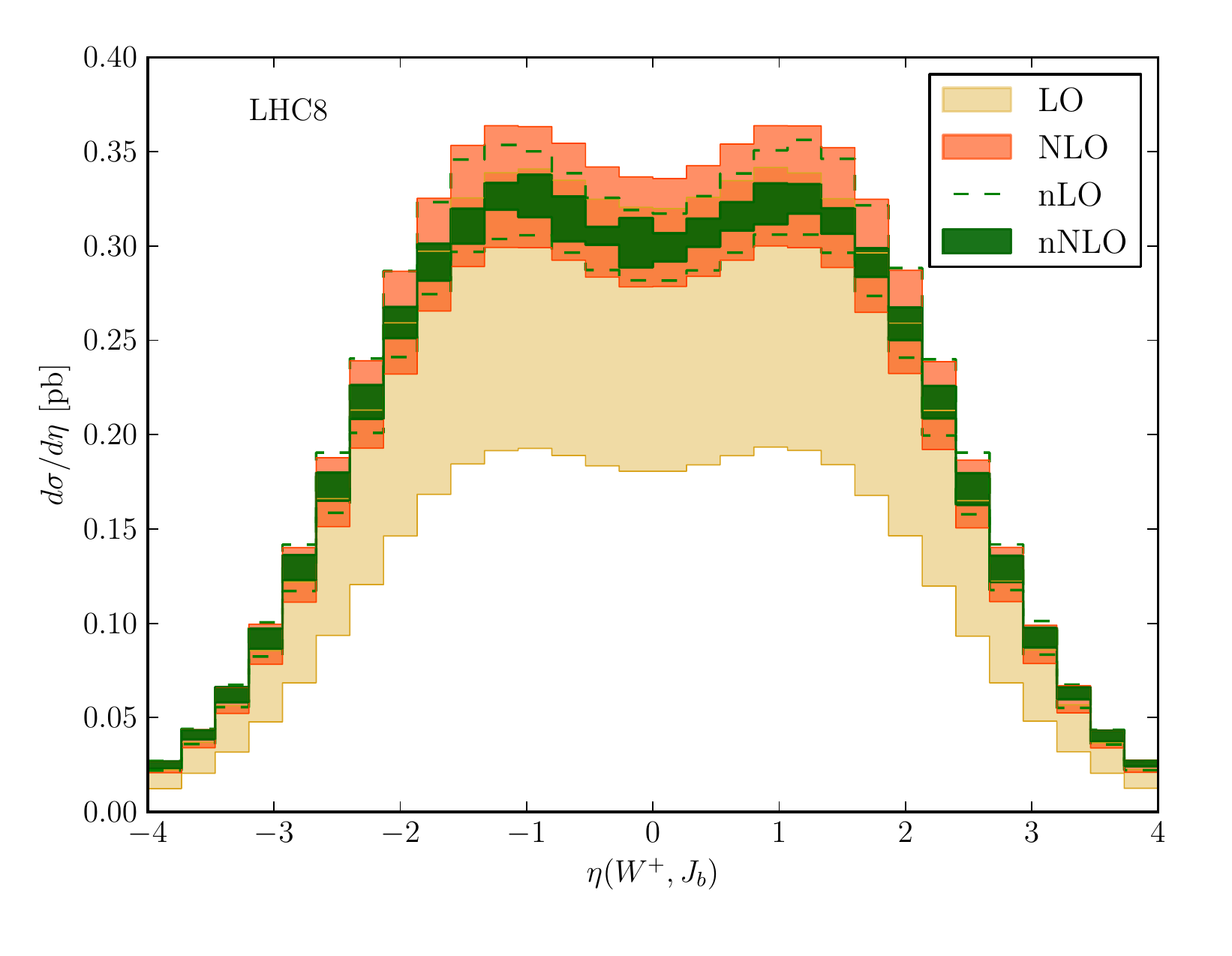}
& \quad
\hspace{-0.6cm}
\includegraphics[trim=0.3cm 0.6cm 0.7cm 0.4cm,clip,width=0.48\textwidth]{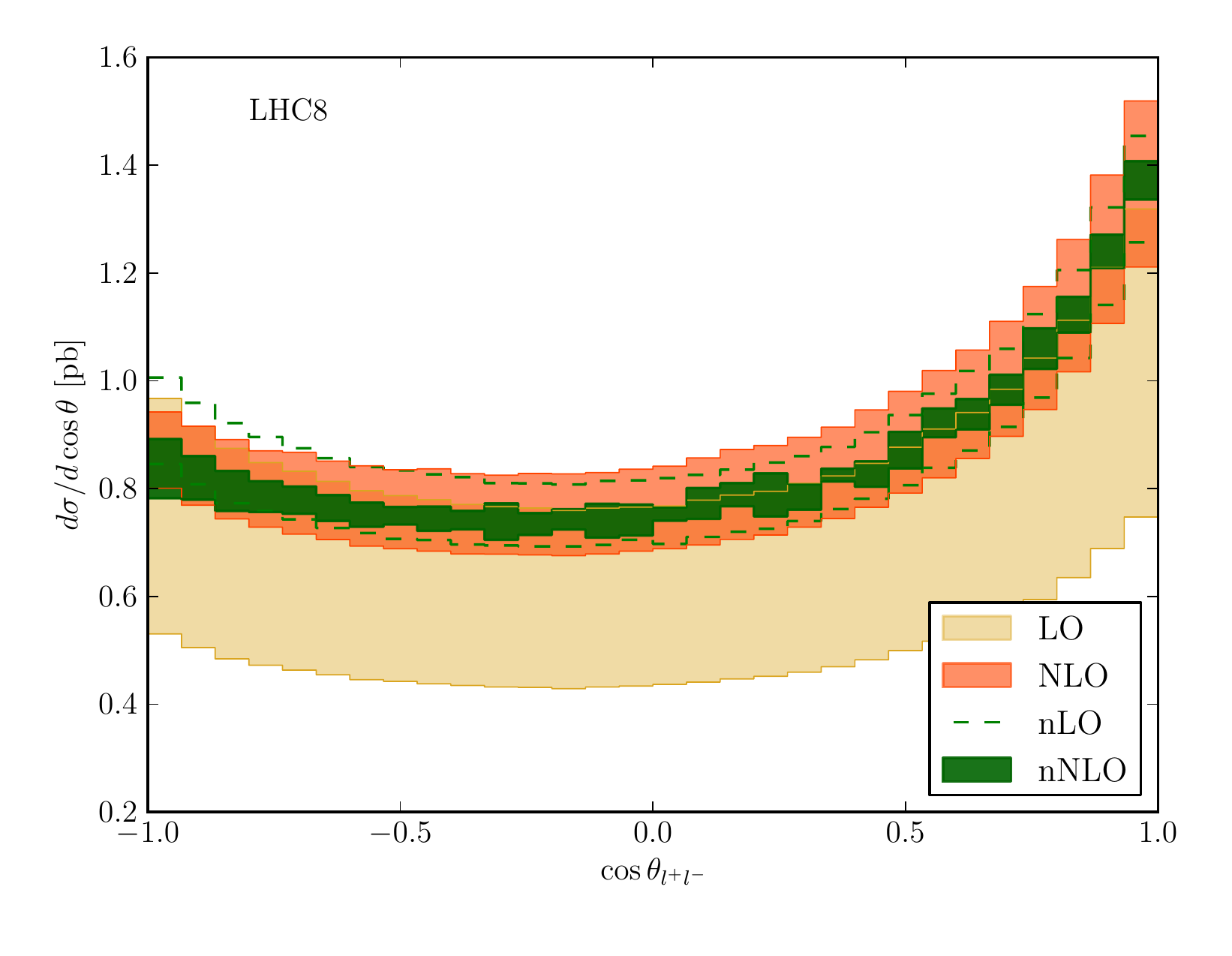} \\ 
\hspace{-0.5cm}
\includegraphics[trim=0.3cm 0.6cm 0.7cm 0.4cm,clip,width=0.48\textwidth]{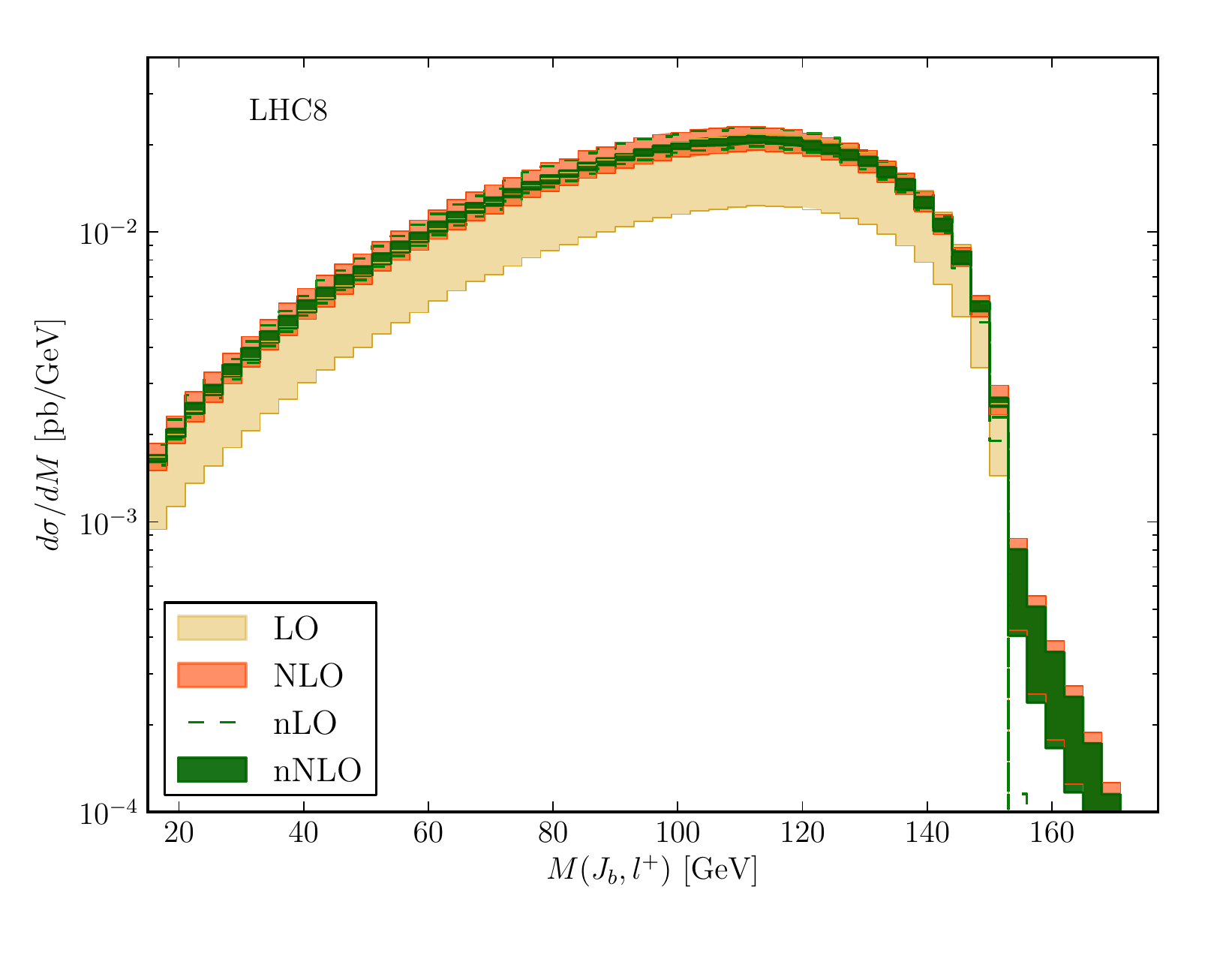} & \quad
\hspace{-0.5cm}
\includegraphics[trim=0.3cm 0.6cm 0.7cm 0.4cm,clip,width=0.48\textwidth]{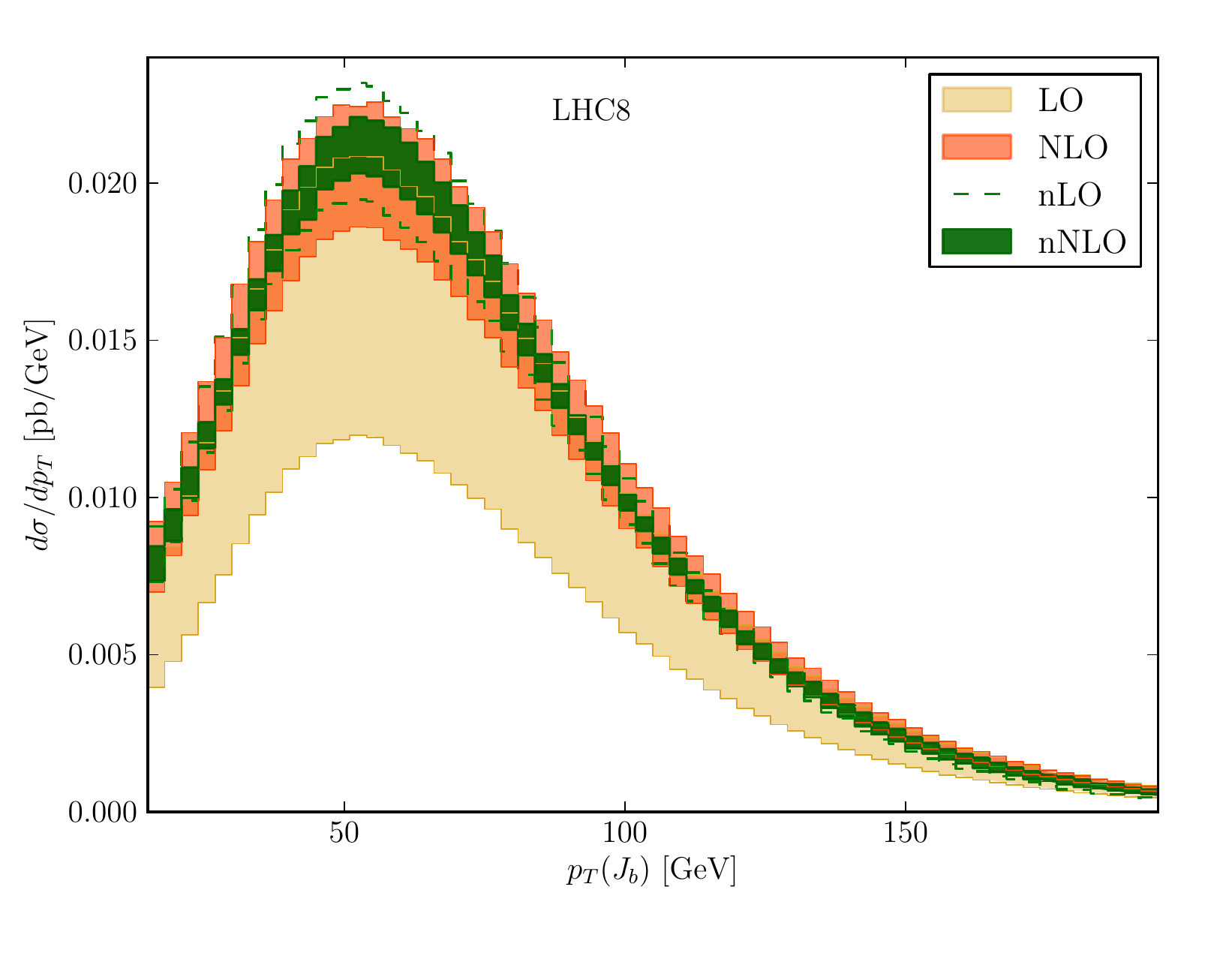} \\ 
\end{tabular}
\caption{Generic observables for LHC8 with analysis cuts. }
\label{fig:lhc8}
\end{figure}

The picture is very much the same for other observables. As an example
we consider the pseudorapidity of the reconstructed top, $\eta(W^+,J_b)$, 
in the top left
panel of Figure~\ref{fig:lhc8}. One would expect that 1PI is the
appropriate kinematics for this observable, but as for the transverse
momentum of the top, PIM kinematics gives very similar results. This
is further evidence that the bulk of the corrections is independent of
the precise details of the kinematics. We can also study generic
observables that do not necessarily have a direct link to either the
PIM or 1PI kinematics. For example, we show in Figure~\ref{fig:lhc8}
the cosine of the angle between the two charged leptons, $\cos\theta_{l^+l^-}$
(top right panel) which is interesting in the study of angular correlations, 
the invariant mass of the lepton-jet system, $M(J_b,l^+)$ (lower left panel) 
which is a useful observable to measure the top mass, and the transverse 
momentum of the $b$-jet, $p_T(J_b)$ (lower right panel). For all these observables the general
features are the same in that the nNLO bands are mostly within the NLO
bands and that the nLO approximation has a very large overlap with the
full NLO band. 

In the upper tail of the invariant mass of the lepton-jet system,
$M(J_b,l^+)$ the nNLO band is of the same size as the NLO band.  This
is as expected given that in the NWA this region of phase space only
receives corrections from (hard) real emission corrections which are
only included via our matching with the NLO. In any case, this region
is not reliably predicted since our results have been obtained with
strictly on-shell top quarks and it is known that close to kinematic
boundaries such as the upper edge of the $M(J_b,l^+)$ distribution,
off-shell effects are sizeable~\cite{Denner:2012yc, Falgari:2013gwa}.

\begin{figure}[t!]
\centering
\begin{tabular}{c c}
\hspace{-0.5cm}
\includegraphics[trim=0.3cm 0.6cm 0.7cm 0.4cm,clip,width=0.48\textwidth]{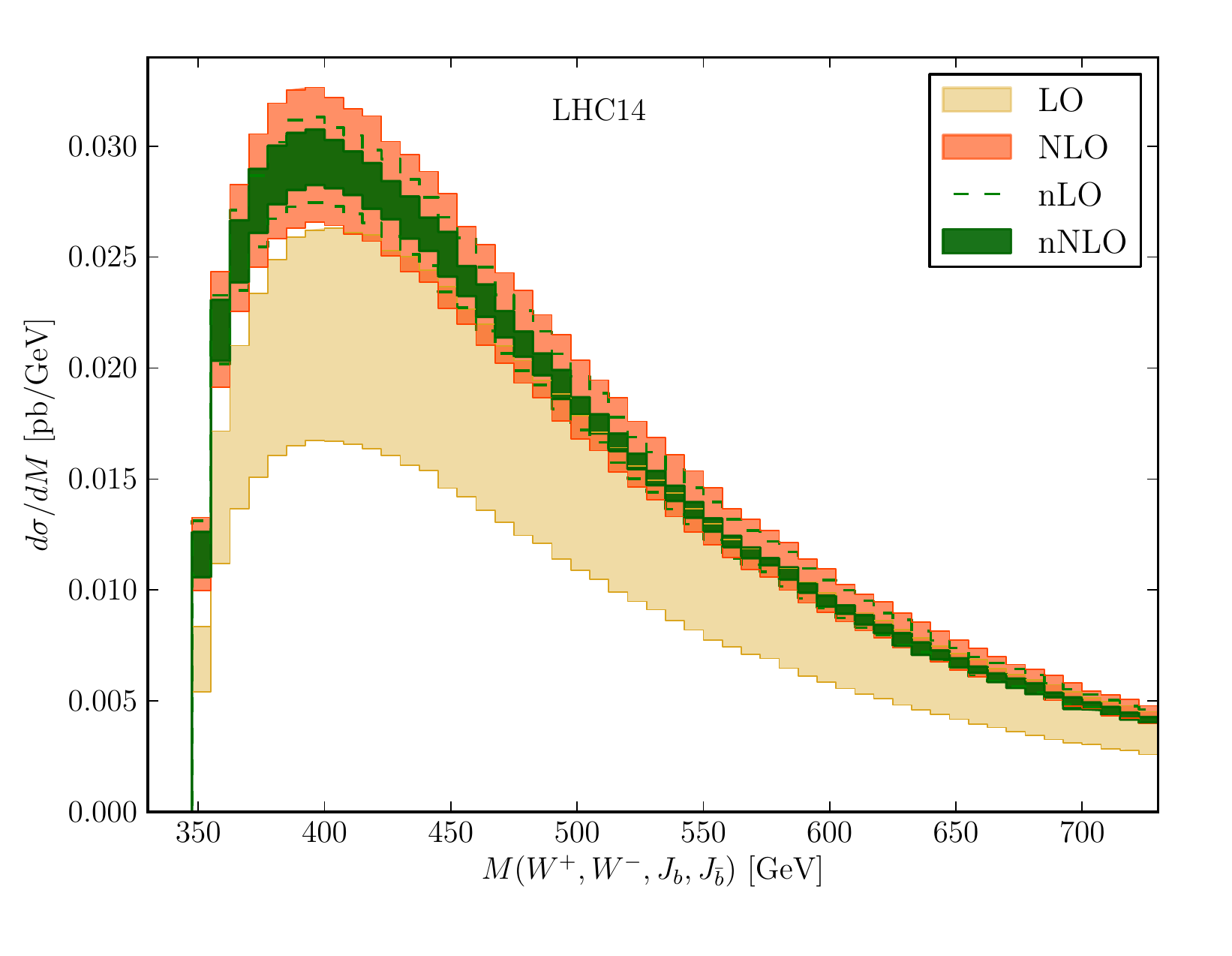} & \quad
\hspace{-0.6cm}
\includegraphics[trim=0.3cm 0.6cm 0.7cm 0.4cm,clip,width=0.48\textwidth]{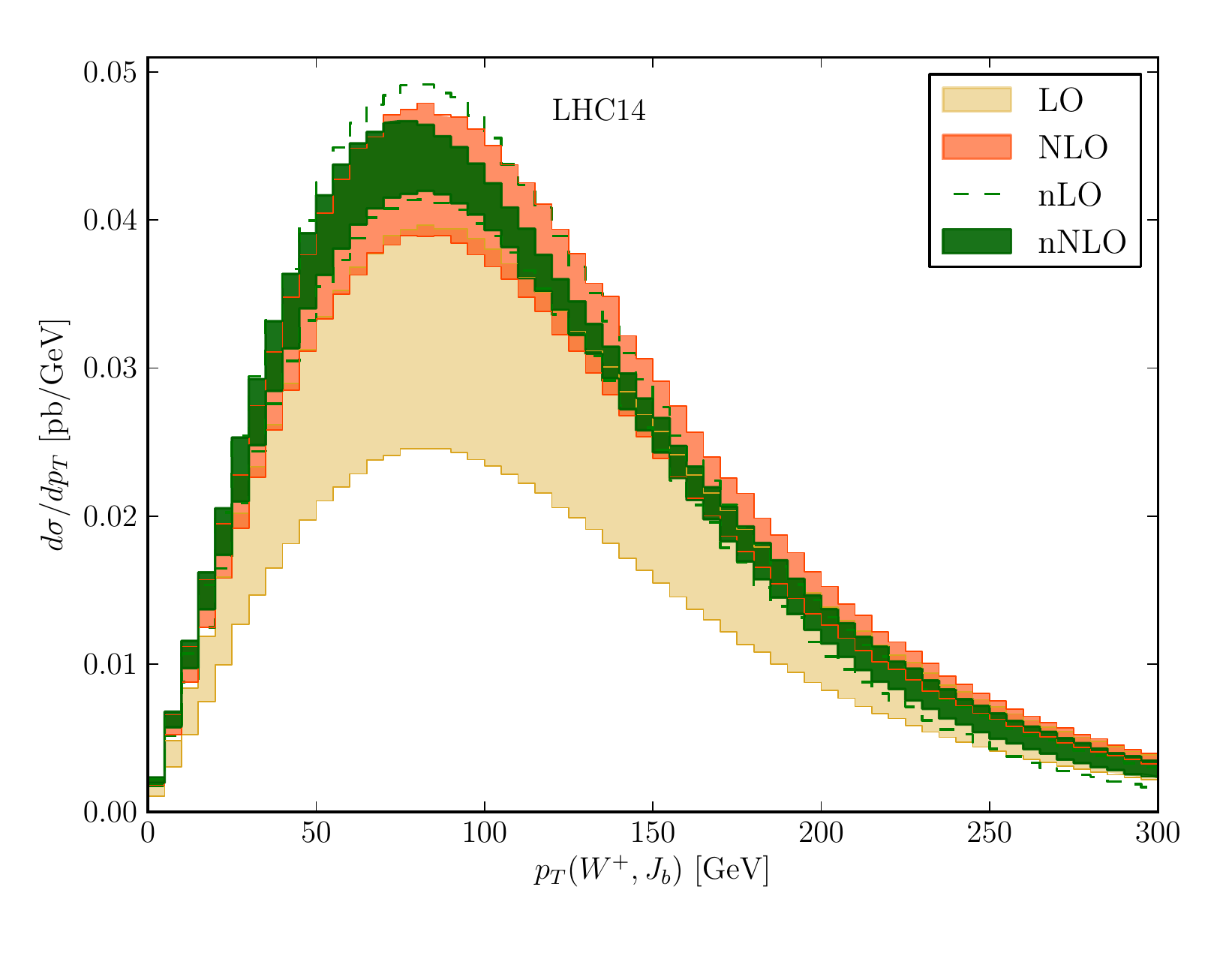} \\ 
\hspace{-0.5cm}
\includegraphics[trim=0.3cm 0.6cm 0.7cm 0.4cm,clip,width=0.48\textwidth]{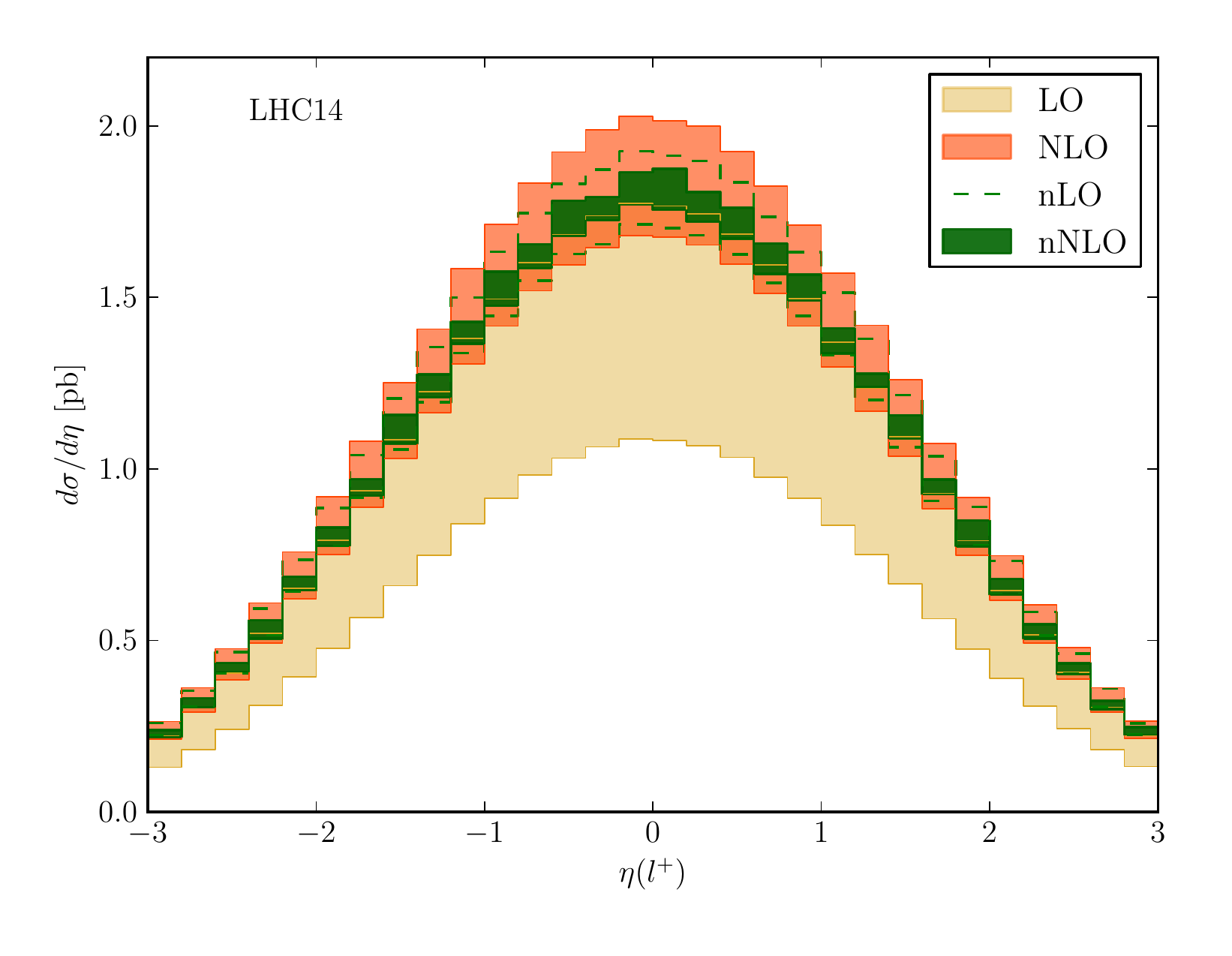} & \quad
\hspace{-0.5cm}
\includegraphics[trim=0.3cm 0.6cm 0.7cm 0.4cm,clip,width=0.48\textwidth]{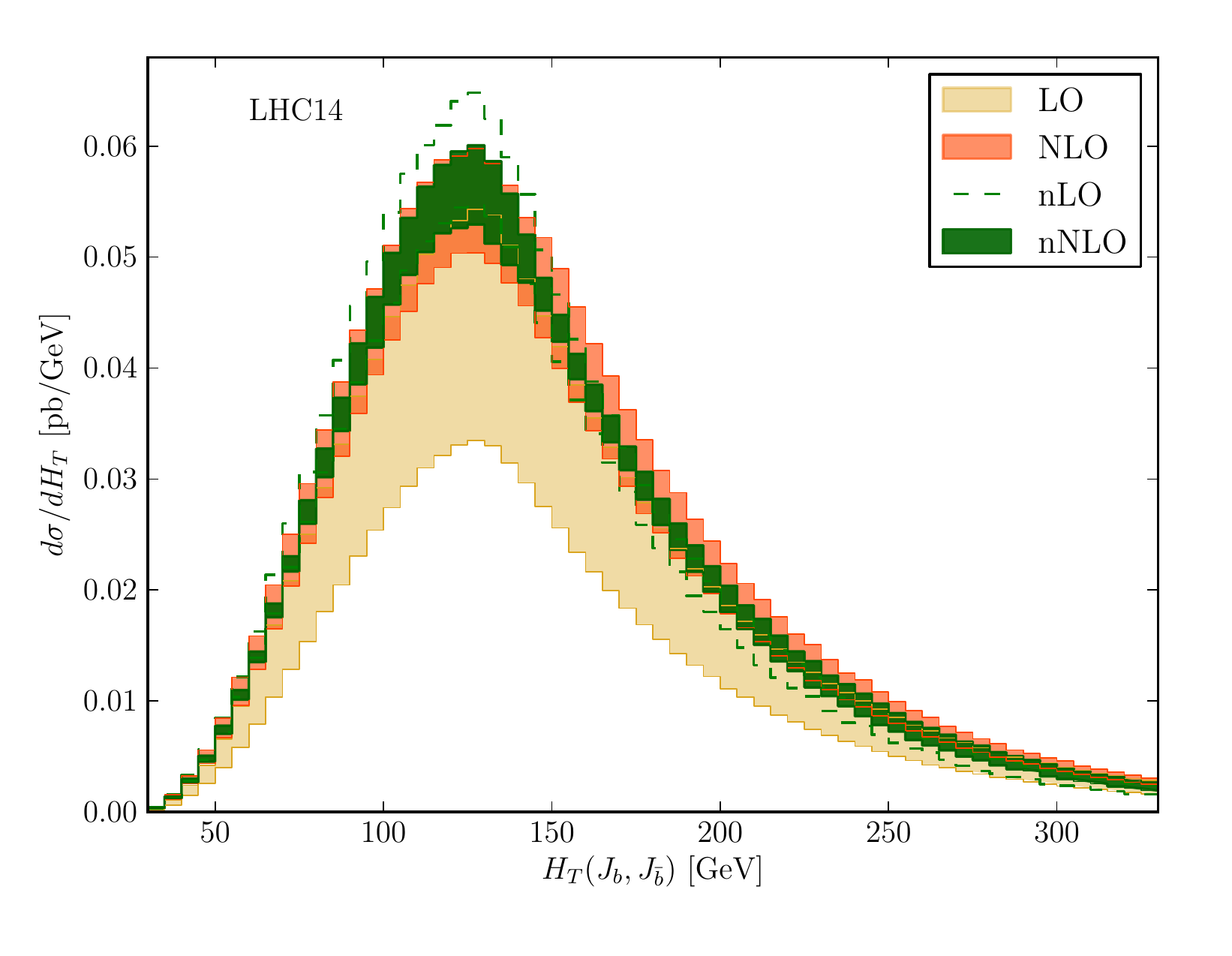} \\ 
\end{tabular}
\caption{Observables for LHC14 with analysis cuts. }
\label{fig:lhc14}
\end{figure}

The main features of the results are not changed at all if we consider
the LHC at the larger energy of 14~TeV. To illustrate this we show
some sample results for LHC14 in Figure~\ref{fig:lhc14} with the same
analysis cuts as for LHC8. In the top two panels we show our standard
observables, $M(W^+,W^-,J_b,J_{\bar{b}})$ and $p_T(W^+,J_b)$.  Again,
applying PIM and 1PI kinematics gives very consistent results for
these two observables apart from the high $p_T$ tail. This justifies
the application of our method to other observables. As an example, in
the lower left panel of Figure~\ref{fig:lhc14} we show $\eta(l^+)$,
the pseudorapidity of the charged lepton from the decay of the
top. Once more we find a very consistent picture when comparing the
various approximations and we have verified that this is that case for
a large number of standard observables. We mention in passing that if
we were to consider observables related to the decay products of the
anti-top rather than the top quark it would be more appropriate to
implement the 1PI kinematics where the $\bar{t}$ is singled out. This
would only introduce a trivial change in the implementation.

Of the many observables we have studied, we now consider one of the
most problematic cases, where our approximation is less reliable: in
the lower right panel of Figure~\ref{fig:lhc14} we consider
$H_T(J_b,J_{\bar{b}})$, the scalar sum of the transverse momenta of
the two $b$-jets. There are notable differences between the nLO and
NLO bands and the reduction in the error band from NLO to nNLO is less
pronounced. This is an observable that combines information from the
decay products of the $t$ and the $\bar{t}$. Hence the applicability
of the 1PI kinematics could be problematic and, indeed, we have
verified that it is the 1PI kinematics that leads to the enhancement
in the peak region and to the undershoot in the tail region of the nLO result
compared to the NLO result. If we were to take the envelope of the PIM
results only, the nLO band would be perfectly consistent with the NLO
band. This is in line with the general observation that the
approximate results using PIM kinematics seem to lead to more
consistent results. In this context we should also stress that
whenever an observable is strongly affected by hard gluon radiation,
our approximate results cannot be trusted.

\begin{figure}[t]
\centering
\begin{tabular}{c c}
\includegraphics[trim=0.2cm 0.6cm 0.6cm 0.4cm,clip,width=0.48\textwidth]{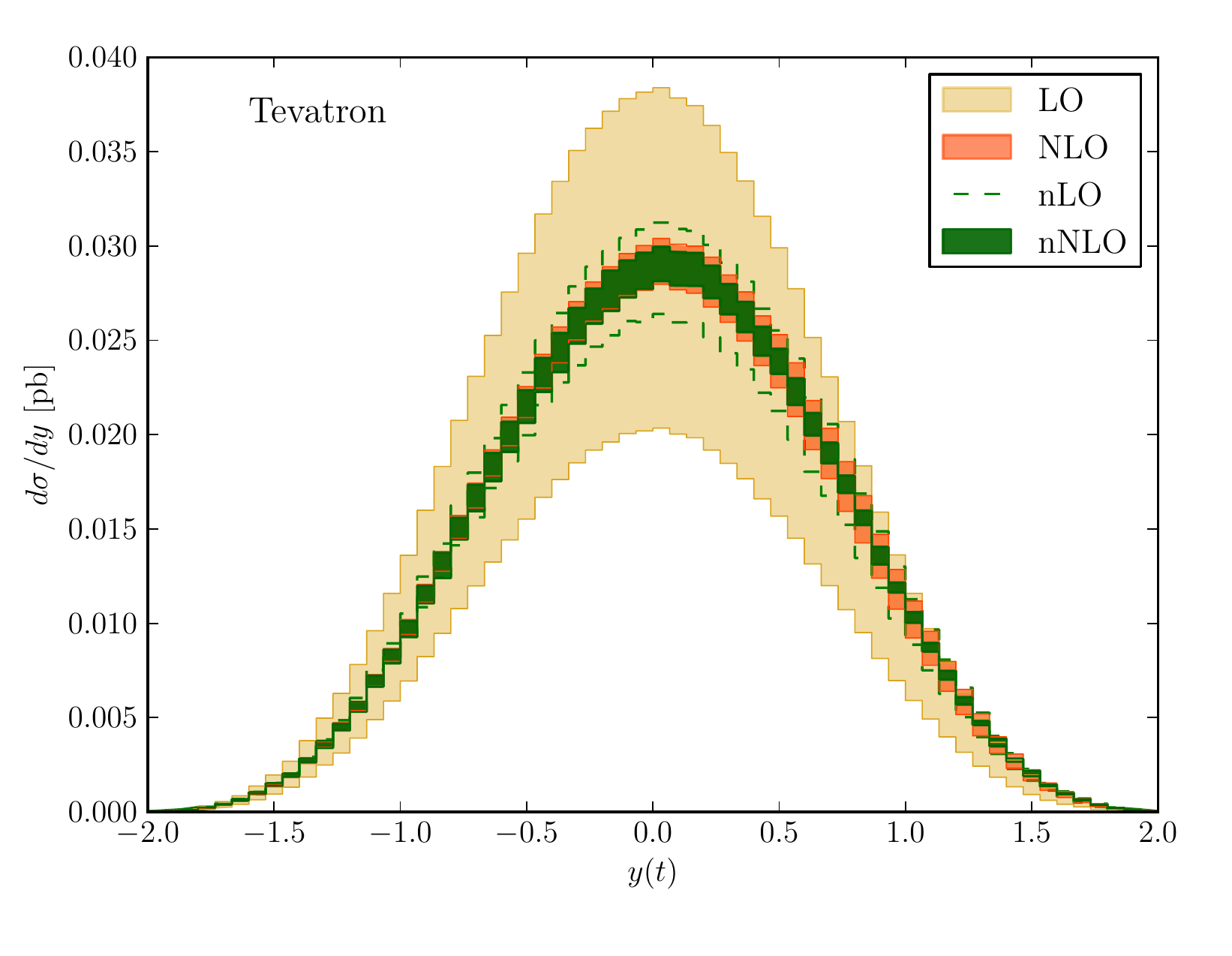} &
\includegraphics[trim=0.2cm 0.6cm 0.6cm 0.4cm,clip,width=0.48\textwidth]{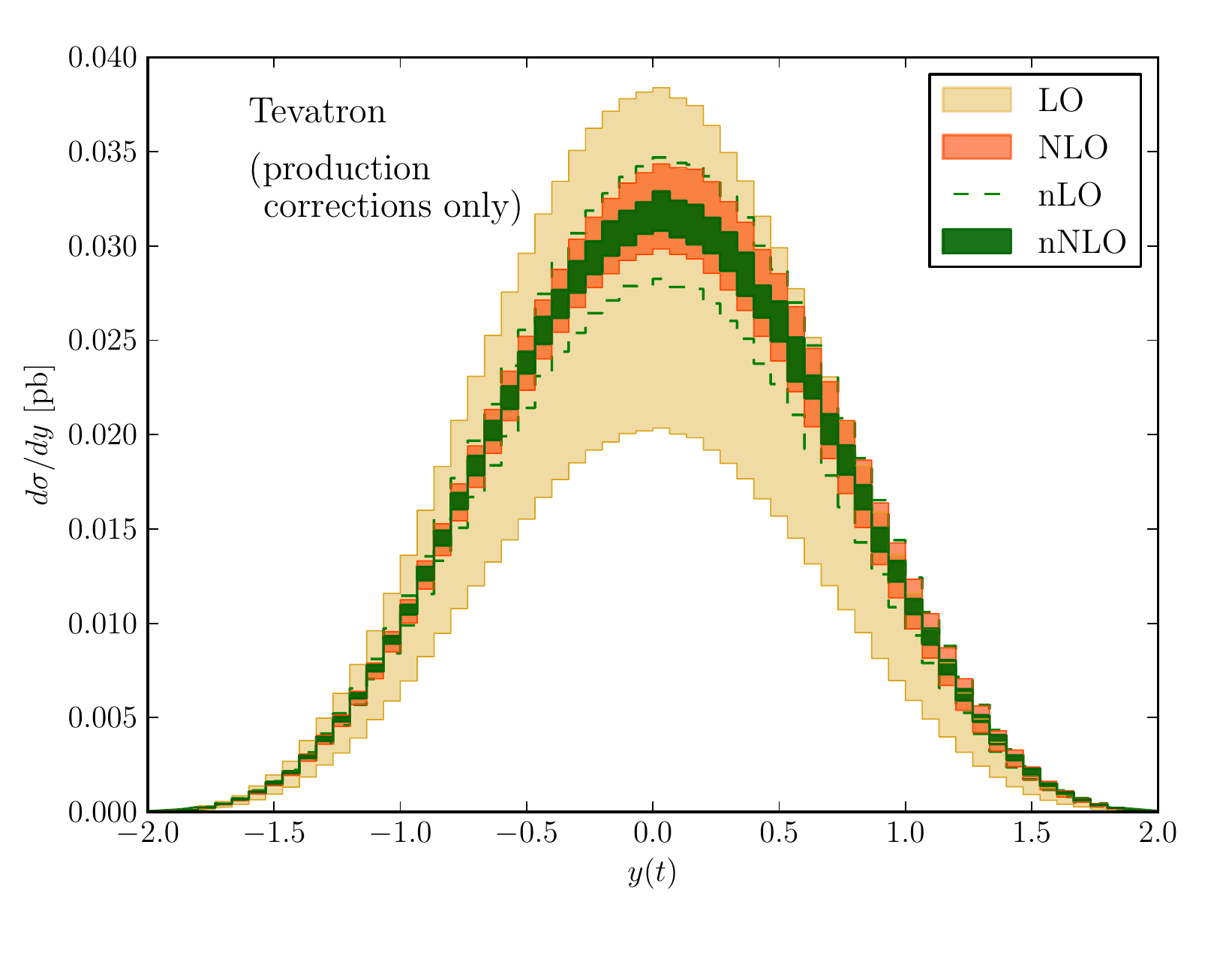}
\end{tabular}
\caption{Rapidity of reconstructed top quark for Tevatron with
  analysis cuts. nNLO corrections to production and NLO corrections to
  decay have been included in left panel, whilst right panel has NLO
  corrections to top decays omitted.}
\label{fig:yt-tev-wcuts-wdec}
\end{figure}

Finally we consider distributions for the Tevatron. We have seen in
Section~\ref{sec:resTotXS} that in the case of the Tevatron the
agreement between NNLO and nNLO for the total cross section is
marginal, casting doubts on the reliability of the approximation we
make. Furthermore, for the cuts we have chosen, the NLO corrections to
the decay are particularly important at the Tevatron. Given this, and
that we use a pure NLO calculation for the decay corrections we are
not expecting any improvement beyond NLO in these circumstances. Let
us illustrate this for the rapidity of the (reconstructed) top
quark. In addition to the LO result, the results including the decay
corrections, $d\sigma^{\text{nLO}}_{\text{full}}$,
$d\sigma^{\text{NLO}}_{\text{full}}$ and
$d\sigma^{\text{nNLO}}_{\text{full}}$ are shown in the left panel of
Figure~\ref{fig:yt-tev-wcuts-wdec}. While
$d\sigma^{\text{nLO}}_{\text{full}}$ is perfectly consistent with
$d\sigma^{\text{NLO}}_{\text{full}}$, there is virtually no
improvement in going from $d\sigma^{\text{NLO}}_{\text{full}}$ to
$d\sigma^{\text{nNLO}}_{\text{full}}$. This is to be contrasted with
the results shown in the right panel of
Figure~\ref{fig:yt-tev-wcuts-wdec}, where only the corrections to the
production have been taken into account. Firstly, comparing the red
bands on the right panel ($d\sigma^{\text{NLO}}_{\text{prod}}$) to the
left panel ($d\sigma^{\text{NLO}}_{\text{full}}$) we see the decisive
impact of the NLO corrections to the decay. This is mainly due to the
cuts imposed on the transverse momenta of the $b$-jets. Secondly, if
only the corrections to the production are taken into account, the
improvement in going from NLO (red band on right panel) to nNLO (green
band on right panel) is apparent.

To sumarize, we have investigated a large number of observables for the 
LHC8, LHC14 and the Tevatron and the results we have found are well 
represented by the examples we have presented in this subsection. While the
applicability of the method in its current implementation for the
Tevatron is questionable, in the case of the LHC our nNLO results for
distributions are very likely to include the bulk of the NNLO
corrections to the production. The advantage of our results is that in
addition to the corrections to the production, the NLO corrections to
the decay are also included.  Depending on the cuts that are applied,
these corrections can be very important. At the same time, we
reiterate that currently we have no improvements beyond NLO for the
decay part.

\section{Future improvements and conclusions}
\label{sec:conclusions}

\subsection{Possible improvements}

While the generic set of differential results we have presented in the
previous section paint a consistent picture as far as perturbative
stability is concerned, there are clear hints that this is not the
full picture and that our results can be improved in a number of ways.

Firstly, given that the nNLO inclusive cross section underestimates
the full NNLO cross section, we are motivated to understand the reason
behind this. Since the nLO cross section approximates the NLO cross
section well, this points towards that reason being, to a large
extent, due to the terms we miss from the two-loop soft and hard
functions at NNLO. Once these become available it is easy to include
them in our setup and we fully expect to see an improvement in the
agreement of nNLO vs NNLO. As discussed in
Section~\ref{subsec:mc-implementation}, the approximate results are
obtained through an integration of the hard scattering kernels over an
approximated phase space. The approximations made are perfectly valid
in the soft limits, however it would be interesting to investigate the
possibility of integrating the hard scattering kernels derived in
\cite{Ahrens:2010zv,Ahrens:2011mw} over the complete phase
space instead of over an approximate one. Doing this could well lead 
to positive contributions moving the nNLO cross section closer to 
that of the NNLO.

Secondly, our exploration of scale variations for several
distributions in Section~\ref{subsec:collider-diff-results} highlight
the importance of higher-order corrections in the decay, when being
exclusive in the top decay products. This is particularly evident in
the Tevatron results of Figure~\ref{fig:yt-tev-wcuts-wdec}.  In order
to obtain more reliable results and in particular a more reliable
uncertainty estimate, it is clear that the accuracy to which the top
decay is computed must ideally match that of the accuracy of the
production subprocess. The NNLO QCD corrections to the top decay have
been recently computed in \cite{Gao:2012ja,Brucherseifer:2013iv} and a
natural realistic next step is to include these in our framework.

Finally, in view of the increasing need to match the experimental
precision on the measurement of the top quark mass parameter with an
equally good theory interpretation, it is important to be able to
describe as realistically as possible the spectra of the top decay
products. In particular this involves relaxing the assumption of the
NWA and treating the top quark as off-shell. This can
straightforwardly be done, while including the nNLO corrections in the
production, as an extension of the effective-theory approach of
\cite{Falgari:2013gwa}.  Furthermore, it is also important to study
the effect on fully differential observables of using a
renormalization scheme other than the pole scheme for the top mass,
something which again is possible within the framework we have
presented here.\footnote{We bring to the reader's attention the recent
  work \cite{Dowling:2013baa} in which the differences between the
  pole and $\overline{\text{MS}}$ schemes have been studied for the
  rapidity and transverse momentum of the top (stable) in which a
  better perturbative convergence is observed in the
  $\overline{\text{MS}}$ scheme.}

\subsection{Conclusions}

In this work we have presented an extension of the approximate (N)NLO
cross section for stable $t\bar{t}$ production, differential in ($M$
and $\cos\theta$) or ($p_T$ and $y$) in PIM and 1PI kinematics
respectively, computed in \cite{Ahrens:2010zv,Ahrens:2011mw}.  The
spin-correlated LO decays of the top quarks have been included in a
new calculation of the hard functions. Our best prediction consists of
the approximate NNLO corrections to the production subprocess, matched
to the full NLO results, where NLO corrections in the decay are also
included in the NWA. We have implemented the approximate (N)NLO
corrections in a fully differential parton-level Monte Carlo that
allows us to study arbitrary (IR-safe) observables constructed out of
the final state jet and lepton momenta.

The flexibility of having the approximate corrections implemented in a
Monte Carlo made it possible to assess how well the approximate NLO
results model the NLO for observables other than the ones they were
designed for. From an examination across colliders and energies and
for an example cut-based analysis, we empirically conclude that the
approximate results for generic observables are largely independent of
the specific soft-gluon kinematics on which they are built. It also
appears that the hard kernels constructed in the PIM kinematics do a
better job of approximating the NLO, even, surprisingly, for the case
of the $p_T$ distribution where one would expect 1PI to do better.

While our results are very encouraging, in that in most cases they
display good perturbative convergence, unfortunately for the inclusive
cross section the approximate NNLO numbers undershoot the full
NNLO. We have argued that the reason is likely due to missing terms in
our approximate approach and also perhaps due to the treatment of
(approximate) phase space. To account for this additional theory error
in our results, we choose to take the envelope of PIM and 1PI scale
uncertainty bands (as opposed to just PIM or just 1PI) and in doing so
expect to find overlap with the NNLO bands once differential results
become available for the latter.

Finally, we reiterate that there is a mismatch between the accuracy to
which the production and decay subprocesses are each computed. This is
reflected in some of the plots shown (or some regions of phase space)
where the scale uncertainty does not decrease when going from NLO to
nNLO. Once again, this underlines the importance of higher order
corrections in the decay, when studying in detail observables
sensitive to the top decay products. The inclusion of NNLO corrections
to the decay, together with possible improvements for the production
part mentioned above, offer a clear path for further improvements in
our approach.

\section{Acknowledgements}

We would like to thank Simon Badger for helpful correspondence on the
one-loop helicity amplitudes for top pair production and Andrea
Ferroglia for very useful discussions.

\newpage
\bibliography{ttbar_xnnlo}
  
\end{document}